\documentclass[12pt,a4]{article}
\usepackage{sw20lart}

\input tcilatex
\begin{document}

\title{\textbf{Algebraic Renormalization \newline}\\
\textit{Perturbative twisted considerations on }\\
\textit{topological Yang-Mills theory and on }\\
\textit{N=2 supersymmetric gauge theories }}
\author{\textbf{F. Fucito, A. Tanzini} \\
Dipartimento di Fisica, Universit\'{a} di Roma II, \\
''\textit{Tor Vergata '', }Italy\\
\vspace{1.5mm} \and \textbf{L.C.Q.Vilar, O. S. Ventura, C.A.G. Sasaki } \\
CBPF, Centro Brasileiro de Pesquisas F\'{\i}sicas \\
Rua Xavier Sigaud 150, 22290-180 Urca \\
Rio de Janeiro, Brazil\vspace{1.5mm}\\
and\vspace{1.5mm} \and \textbf{S.P. Sorella } \\
UERJ, Universidade do Estado do Rio de Janeiro\\
Departamento de F\'{\i}sica Te\'{o}rica, Instituto de F\'{\i}sica\\
Rua S\~ao Francisco Xavier, 524\\
20550-013, Maracan\~{a}, Rio de Janeiro, Brazil}
\maketitle

\begin{abstract}
Lectures given at the ''\textit{First School on Field Theory and Gravitation
'', }Vit\'{o}ria, Esp\'{\i}rito Santo, Brazil, 15-19 April, 1997.

\setcounter{page}{0}\thispagestyle{empty}
\end{abstract}

\vfill\newpage\ \makeatother
\renewcommand{\theequation}{\thesection.\arabic{equation}}

\section{\ Introduction\-}

The aim of these notes is to provide a simple and pedagogical (as much as
possible) introduction to what is nowadays commonly called \textit{Algebraic
Renormalization }\cite{book}. As the name itself let it understand, the 
\textit{Algebraic Renormalization} gives a systematic set up in order to
analyse the quantum extension of a given set of classical symmetries. The
framework is purely algebraic, yielding a complete characterization of all
possible anomalies and invariant counterterms without making use of any
explicit computation of the Feynman diagrams. This goal is achieved by
collecting, with the introduction of suitable ghost fields, all the
symmetries into a unique operation summarized by a generalized
Slavnov-Taylor (or master equation) identity which is the starting point for
the quantum analysis. The Slavnov-Taylor identity allows to define a
nilpotent operator whose cohomology classes in the space of the integrated
local polynomials in the fields and their derivatives with dimensions
bounded by power counting give all nontrivial anomalies and counterterms. In
other words, the proof of the renormalizability is reduced to the
computation of some cohomology classes.

However, before going any further, let us make some necessary remarks on the
limitations of the method. The \textit{Algebraic Renormalization }applies
basically to the perturbative regime, meaning that the quantum extension of
the theory is constructed order by order in the loop parameter expansion $%
\hbar $. Aspects regarding the convergence and the resummation of the
perturbative series are not considered and cannot, in general\footnote{%
With the exception of few two dimensional models.}, be handled within this
algebraic setup. In spite of its perturbative character the \textit{%
Algebraic Renormalization}, being based only on the locality and power
counting properties of the renormalization theory, does not rely on the
existence of any regularization preserving the symmetries. This means that
the algebraic proofs of the renormalizability extend to all orders of
perturbation theory and are independent from the regularization scheme. This
very important feature gives to the \textit{Algebraic Renormalization} a
very large domain of applicability. In practice, one can include almost all
known power counting renormalizable theories in flat space-time covering, in
particular, those for which no invariant regularization is known. It is also
worthwhile to mention that, besides the pure characterization of the
anomalies and of the invariant counterterms, the \textit{Algebraic
Renormalization }plays a quite useful role in the study of other aspects of
field theory which, although not touched in these lectures, are object of a
fruitful research activity. Let us mention, for instance:

\vspace{5mm}

$\bullet $\ \ \textbf{The nonrenormalization theorems and the ultraviolet
finiteness }

\textit{The aim here is to establish nonrenormalization theorems for
anomalies (as the Adler-Bardeen nonrenormalization theorem of the gauge
anomaly \cite{adba}) and to provide a classification of models which have
vanishing }$\beta $-\textit{function to all orders of perturbation theory.
Examples of the latters are given by some four dimensional gauge theories
with }$N=1,2,4$\textit{\ supersymmetry \cite{oliv,white} and by some
topological field theories \cite{tf,tft}.}

\vspace{5mm}

$\bullet \;\;$\textbf{The characterization of new symmetries}

\textit{This aspect, deeply related to the previous one, consists of the
search of additional unknown symmetries (eventually linearly broken) which
may be responsible for the finiteness properties displayed by a particular
model. This problem, related to the existence of cohomology classes\footnote{%
It turns out that in some cases also certain exact cocycles with negative
ghost number become relevant \textit{\cite{vsusy}}.} with negative ghost
number, allows for a cohomological (re)interpretation of the Noether theorem 
\cite{noether}. Examples of such additional symmetries are provided by the
so called vector supersymmetry of the topological theories and by the Landau
ghost equation responsible for the ultraviolet finiteness of a wide class of
invariant local field polynomials in the Yang-Mills theories \cite{vsusy}.}

\vspace{5mm}

$\bullet $\ \ \textbf{The geometrical aspects}

\textit{As one can easily expect, the number of applications related to the
geometrical aspects of the BRST transformations and of the anomalies is very
large. We shall limit ourselves only to mention a particular feature which
is focusing our attention since a few years, namely the possibility of
encoding all the transformations of all the fields and antifields (or BRST
external sources) into a unique equation which takes the form of a
generalized zero curvature condition \cite{zerocurv}. The zero curvature
formalism allows us to obtain in a very simple way the BRST\ cohomology
classes and improves our understanding of the role of the antifileds.}

\vspace{5mm}

Having (hopefully) motivated the usefulness of the \textit{Algebraic
Renormalization, }let us now briefly describe the plan of these notes. We
shall adopt here the point of view of not entering into the technical
aspects concerning the antifields formulation and the computation of the
BRST cohomology classes, limiting ourselves only to state the main results
and reminding the reader to the several reviews and books appeared recently
in the literature \textit{\cite{book,ht}}. Rather, we shall work out in
detail the renormalization of a model rich enough to cover all the main
aspects of the algebraic method. The example we will refer to is the four
dimensional euclidean topological Yang-Mills (TYM) theory proposed by E.
Witten \textit{\cite{tym} }at the end of the eighty's. Besides the mere fact
that TYM is a continuous source of investigations, we shall see that this
model possesses a very interesting structure, requiring a highly nontrivial
quantization and displaying peculiar cohomological properties. This is due
to the deep relationship with the $N=2$ euclidean supersymmetric Yang-Mills
theory. In fact TYM in flat space-time can be actually seen as the twisted
version of the $N=2\;$Yang-Mills theory in the Wess-Zumino gauge \textit{%
\cite{mar}}, the Witten's fermionic symmetry being identified with the
singlet generator of the twisted $N=2$ supersymmetric algebra. Furthermore,
by means of the introduction of appropriate constant ghosts associated to
the twisted $N=2$ generators, we shall be able to quantize the model by
taking into account both the gauge invariance and the $N=2$ supersymmetry,
overcoming the well known difficulties of the $N=2$ susy algebra in the
Wess-Zumino gauge \textit{\cite{white, magg}}. Concerning now the BRST
cohomology, we will have the opportunity of checking how the twisted $N=2\;$%
susy algebra can be used to obtain in a straightforward way the relevant
cohomology classes. In particular, it will turn out that the origin of the
TYM\ action can be traced back to the invariant polynomial $tr\left( \phi
^2\right) $, $\phi $ being one of the scalar fields of the model. This
relation has a very appealing meaning. Needless to say, the $N=2$ susy YM
theory is indeed the corner stone of the duality mechanism recently
discussed by N. Seiberg and E. Witten \textit{\cite{sw}}, who used in fact $%
tr\left( \phi ^2\right) $\ in order to label the different vacua of the
theory. Finally, we will show that the requirement of analyticity \textit{%
\cite{bc} }in the constant ghosts of the twisted $N=2\;$supersymmetry can be
deeply related to the so called \textit{equivariant} \textit{cohomology}
proposed by R. Stora et al.\textit{\cite{stora1,stora2} }in order to deal
with the topological theories of the cohomological type.

\newpage\ 

\section{Generalities on the Slavnov-Taylor identity and on Cohomology}

\subsection{Classical action and symmetry content}

The starting point of our analysis consists of assigning a set of fields $%
(A_\mu ,\left\{ \lambda \right\} )$, $A_\mu $\ and $\left\{ \lambda \right\}
\;$being respectively a gauge field and a set of spinor and scalar matter
fields, and a classical gauge invariant action $S_{inv}$

\begin{eqnarray}
S_{inv} &=&\int d^4x\mathcal{L(}A,\lambda )\;,\;  \label{inv-action} \\
\delta _\epsilon ^gS_{inv} &=&0\;,  \nonumber
\end{eqnarray}
with

\begin{eqnarray}
\delta _\epsilon ^gA_\mu &=&-D_\mu \epsilon =-\left( \partial _\mu \epsilon
+[A_\mu ,\epsilon ]\right) \;,  \label{gauge-transf} \\
\delta _\epsilon ^g\lambda &=&[\epsilon ,\lambda ]\;,  \nonumber
\end{eqnarray}
where $\mathcal{L(}A,\lambda )\;$is a power counting renormalizable local
polynomial in the fields and their derivatives and $\delta _\epsilon ^g\;$is
the generator of the gauge transformations with local infinitesimal
parameter $\epsilon (x)$. All the fields are Lie algebra valued, \textit{i.e.%
} $A_\mu =A_\mu ^aT^a\;$and $\lambda =\lambda ^aT^a$, the generators of the
corresponding gauge group $G$ being chosen to be antihermitians $\left[
T^a,T^b\right] =f_{\;\;\;c}^{ab}T^c$, with\ $f_{\;\;\;c}^{ab}$ the structure
constants.

\begin{remark}
\textit{The action }$\left( \ref{inv-action}\right) $\textit{\ refers to the
standard case of a model containing only gauge and matter fields, implying
in particular that the only degeneracy in order to compute the propagators
is the one associated with the transverse quadratic term in the gauge fields
following from the usual Yang-Mills term }$tr(F_{\mu \nu }F^{\mu \nu })$. 
\textit{Although} \textit{the algebraic method can be applied to more
sophisticated cases (p-forms,...), the field content of the expression }$%
\left( \ref{inv-action}\right) \;$\textit{covers a very large class of
models, including the TYM\ theory.}
\end{remark}

Besides the gauge invariance, the classical action $\left( \ref{inv-action}%
\right) \;$will be assumed to be left invariant by a set of additional
global transformations whose corresponding generators $\left\{ \delta
_A,\;A=1,...\right\} \;$

\begin{equation}
\delta _AS_{inv}=0\;,  \label{add-inv}
\end{equation}
give rise, together with the gauge generator $\delta _\epsilon ^g,\;$to the
following algebraic relations

\begin{eqnarray}
\left[ \delta _A,\delta _B\right] &=&-C_{\;AB}^C\delta _C+(\mathrm{%
matter\;eqs.\;of\;motion})\;+(\mathrm{gauge\;transf.})\;,
\label{add-algebra} \\
\left[ \delta _A,\delta _\epsilon ^g\right] &=&0\;,  \nonumber
\end{eqnarray}
where $[\cdot ,\cdot ]$\ denotes the graded commutator and \ $C_{\;AB}^C$
are\ appropriate constant coefficients. We do not specify further the nature
of the indices $A$, $B$ of the generators $\delta _A$ which, according to
the particular model considered, may refer to spinor indices, to Lorentz
indices, to group indices, etc... The generators $\delta _A\;$may act
nonlinearly on the fields. The fact that they are related to global
invariances means that, unlike the gauge parameter $\epsilon (x)$\ of eq.$%
\left( \ref{gauge-transf}\right) $, the corresponding infinitesimal
parameters entering the $\delta _A$-transformations do not depend on
space-time. The algebraic structure $\left( \ref{add-algebra}\right) \;$is
typical of supersymmetric gauge theories in the Wess-Zumino gauge \textit{%
\cite{white,magg,dmpw,glob} }and of many topological theories including, in
particular, TYM.

The first step towards the construction of a classical Slavnov-Taylor
identity consists of turning the infinitesimal parameters associated to the
generators $(\delta _\epsilon ^g,\;\delta _A)$\ into suitable ghosts. The
local gauge parameter $\epsilon (x)\;$will be thus replaced by the
Faddeev-Popov ghost $c(x)$\ and $\delta _\epsilon ^g\;$will give rise to the
well known nilpotent operator $s$\ corresponding to the gauge transformations

\begin{eqnarray}
sA_\mu &=&-D_\mu c\;,  \label{brs-tr} \\
s\lambda &=&[c,\lambda ]\;,  \nonumber \\
sc &=&c^2\;,  \nonumber \\
s^2 &=&0\;.  \nonumber
\end{eqnarray}
Concerning now the infinitesimal parameters associated to the $\delta _A$
's, they will be replaced by global constant ghosts $\varepsilon ^A\;$which
will be taken as commuting or anticommuting according to the bosonic or
fermionic character of the corresponding generator. In addition, it can be
shown \textit{\cite{white,magg,dmpw,glob}} that one may define the action of 
$s$\ and of the $\delta _A$'s on the Faddeev-Popov ghost $c$ and on the
global ghosts $\varepsilon ^A\;$in such a way that the extendend BRST
operator given by

\begin{equation}
Q:=s+\varepsilon ^A\delta _A+\frac 12C_{\;AB}^C\varepsilon ^A\varepsilon
^B\frac \partial {\partial \varepsilon ^C}\;,  \label{ext-brst-op}
\end{equation}
has ghost number one and enjoys the following important property

\begin{equation}
Q^2=0\;\;\;\;(\func{mod}.\mathrm{\;matter\;eqs.\;of\;motion)\;}.
\label{q-nilp}
\end{equation}
Of course,

\begin{equation}
QS_{inv}=0\;,  \label{q-inv}
\end{equation}
showing that the operator $Q\;$collects together all the symmetries of the
action $\left( \ref{inv-action}\right) .$

The second step in the construction of the Slavnov-Taylor identity is the
introduction of a gauge fixing term $S_{gf}.$\ To this purpose we introduce
an antighost $\overline{c}\;$and\ a Lagrangian multiplier $b$\ transforming
as

\begin{eqnarray}
Q\overline{c} &=&b+\mathrm{(}\varepsilon -\text{\textrm{dependent terms)\ }},
\label{antigh-tr} \\
Qb &=&0+\mathrm{(}\varepsilon -\text{\textrm{dependent terms)\ }},  \nonumber
\end{eqnarray}
where the $\varepsilon -$dependent terms are chosen in such a way that

\begin{equation}
Q^2\overline{c}=Q^2b=0\;.  \label{antigh-nilp}
\end{equation}
Therefore, recalling from eq.$\left( \ref{q-nilp}\right) \;$that\ $Q^2A_\mu
=0$, it follows that a linear covariant Landau type gauge fixing term

\begin{eqnarray}
S_{gf} &=&Q\int d^4x\;tr(\overline{c}\partial A)  \label{gauge-fix} \\
&=&tr\int d^4x\;\left( b\partial A+\overline{c}\partial ^\mu D_\mu
c+(\varepsilon -\text{\textrm{dep. terms}})\right) \;,
\end{eqnarray}
provides a gauge fixed action $(S_{inv}+S_{gf})$ which is invariant under $Q$%
, \textit{i.e.}

\begin{equation}
Q(S_{inv}+S_{gf})=0\;.  \label{gaug-fixed-inv}
\end{equation}
The above equation means that the gauge fixing procedure has been carried
out in a way which is compatible with all the additional global symmetries $%
\delta _A$\ of the classical action $\left( \ref{inv-action}\right) $. We
also remark that the expression $\left( \ref{gauge-fix}\right) $ belongs to
a class of linear covariant gauge conditions which can be proven to be
renormalizable \textit{\cite{book,magg}}.

We are now ready to write down the Slavnov-Taylor identity. Denoting with $%
\left\{ \varphi ^i\right\} =(A_\mu ,\left\{ \lambda \right\} ,c,\overline{c}%
,b)\;$all the local fields of $(S_{inv}+S_{gf}),$ we associate to each field 
$\varphi ^i\;$of ghost number $N_{\varphi ^i}\;$and\ dimension $d_{\varphi
^i}\;$the corresponding antifield $\varphi ^{*i}\;$with ghost number $%
-(1+N_{\varphi ^i})\;$and dimension $(4-d_{\varphi ^i})$, and we introduce
the antifield dependent action

\begin{equation}
S_{ext}=tr\int d^4x\left( \varphi ^{*i}Q\varphi ^i+\omega _{ij}\varphi
^{*i}\varphi ^{*j}\right) \;.  \label{ext-act}
\end{equation}
The first term in the expression $\left( \ref{ext-act}\right) $ is needed in
order to define the nonlinear $Q$-variations of the fields $\varphi ^i\,$as
composite operators. The second term, quadratic in the antifields, allows to
take care of the fact that the extended operator $Q\,$ of eq.$\left( \ref
{ext-brst-op}\right) $ is nilpotent only modulo the matter equations of
motion. The coefficients $\omega _{ij}$, depending in general from both
fields $\varphi ^i\,$and global ghosts $\varepsilon ^A$, are fixed by
requiring that the following identity holds

\begin{equation}
\mathcal{S(}\Sigma \mathcal{)=\;}0\;,  \label{class-s-t}
\end{equation}
with 
\begin{equation}
\mathcal{S(}\Sigma \mathcal{)=}\int d^4x\frac{\delta \Sigma }{\delta \varphi
^i}\frac{\delta \Sigma }{\delta \varphi ^{*i}}\;+\;\frac
12C_{\;AB}^C\varepsilon ^A\varepsilon ^B\frac{\partial \Sigma }{\partial
\varepsilon ^C}\;,  \label{s-t-op}
\end{equation}
and $\Sigma $ being the complete action

\begin{equation}
\Sigma =S_{inv}\;+\;S_{gf}\;+\;S_{ext}\;.  \label{comp-act}
\end{equation}
The equation $\left( \ref{class-s-t}\right) $\ is called the Slavnov-Taylor
identity and will be the starting point for the quantum analysis.

\begin{remark}
\textit{Although higher order terms (cubic, etc.,..) in the antifields }$%
\varphi ^{*i}\;$\textit{may be required in the action} $\left( \ref{ext-act}%
\right) $\ \textit{in order to obtain the Slavnov-Taylor identity, they will
not be needed in the example considered here.}
\end{remark}

\subsection{Cohomology and renormalizability: anomalies and stability of the
classical action}

We face now the problem of the quantum extension of the classical
Slavnov-Taylor identity $\left( \ref{class-s-t}\right) $, \textit{i.e. }of
the perturbative construction of a renormalized vertex functional\footnote{$%
\Gamma $\ is the generator of the 1PI Green's functions.} $\Gamma $

\begin{equation}
\Gamma =\Sigma +O(\hbar )\;,  \label{1pi}
\end{equation}
fulfilling the quantum version of eq.$\left( \ref{class-s-t}\right) $, 
\textit{i.e.}

\begin{equation}
\mathcal{S(}\Gamma \mathcal{)=\;}0\;,  \label{q-s-t}
\end{equation}
which would imply that all the classical symmetries, \textit{i.e. }the gauge
and the $\delta _A-$invariances, can be implemented at the quantum level
without anomalies. In order to detect the presence of possible anomalies,
let us suppose that eq.$\left( \ref{q-s-t}\right) \;$breaks down at a
certain order $\hbar ^n,\;(n\geq 1),\;$

\begin{equation}
\mathcal{S(}\Gamma \mathcal{)=}\hbar ^n\Delta +O(\hbar ^{n+1})\;,
\label{broken-q-s-t}
\end{equation}
where, from the power counting and locality properties of the renormalized
perturbation theory, the breaking $\Delta \;$is an integrated local
polynomial in the fields, antifields and global ghosts with ghost number one.

The breaking $\Delta \;$is\ easily seen to be constrained by a consistency
condition. In fact, defining the linearized operator $\mathcal{B}_{\mathcal{F%
}}$\ 

\begin{equation}
\mathcal{B}_{\mathcal{F}}\ =\int d^4x\left( \frac{\delta \mathcal{F}}{\delta
\varphi ^i}\frac \delta {\delta \varphi ^{*i}}\;+\frac{\delta \mathcal{F}}{%
\delta \varphi ^{*i}}\frac \delta {\delta \varphi ^i}\right) \;+\;\frac
12C_{\;AB}^C\varepsilon ^A\varepsilon ^B\frac \partial {\partial \varepsilon
^C}\;,  \label{lin-op}
\end{equation}
$\mathcal{F}$ being an arbitrary functional with even ghost number, we have
the following exact algebraic relation

\begin{equation}
\mathcal{B}_{\mathcal{F}}\ \mathcal{S(F)=}\;0\;.  \label{alg-rel}
\end{equation}
In addition, if $\mathcal{F}$ satisfies the Slavnov-Taylor identity 
\begin{equation}
\mathcal{S(F)=\;}0\;,  \label{f-s-t}
\end{equation}
it follows that the operator $\mathcal{B}_{\mathcal{F}}$ \thinspace is
nilpotent,

\begin{equation}
\mathcal{B}_{\mathcal{F}}\mathcal{B}_{\mathcal{F}}=0\;.  \label{f-lin}
\end{equation}
In particular, from the classical Slavnov-Taylor identity $\left( \ref
{class-s-t}\right) \;$it follows that the linearized operator $\mathcal{B}%
_\Sigma \;$corresponding to the classical action $\Sigma $ is nilpotent

\begin{eqnarray}
\mathcal{B}_\Sigma &=&\int d^4x\left( \frac{\delta \Sigma }{\delta \varphi ^i%
}\frac \delta {\delta \varphi ^{*i}}\;+\frac{\delta \Sigma }{\delta \varphi
^{*i}}\frac \delta {\delta \varphi ^i}\right) \;+\;\frac
12C_{\;AB}^C\varepsilon ^A\varepsilon ^B\frac \partial {\partial \varepsilon
^C}\;,  \label{b-lin} \\
\mathcal{B}_\Sigma \mathcal{B}_\Sigma &=&0\;.  \nonumber
\end{eqnarray}

\begin{definition}
\textit{The cohomology classes of the operator }$\mathcal{B}_\Sigma \;$%
\textit{in the space of the local integrated polynomials in the fields }$%
\varphi ^i$\textit{, antifields }$\varphi ^{*i}$\textit{,\ global ghosts }$%
\varepsilon ^A$ \textit{\thinspace and their space-time derivatives, are
defined as the solutions }$\Xi $\textit{\ of the consistency condition }
\end{definition}

\begin{equation}
\mathcal{B}_\Sigma \Xi =0\,\;,  \label{def-coh}
\end{equation}
\textit{which are not of the form }

\begin{equation}
\Xi =\mathcal{B}_\Sigma \widehat{\Xi }\;,\;  \label{triv-coh}
\end{equation}
\ \textit{with }$\left( \Xi ,\widehat{\Xi }\right) \;$\textit{local
integrated polynomials in the fields, antifields and global ghosts.
Solutions of the type\ }$\left( \ref{triv-coh}\right) \;$\textit{are called
exact and can be proven to be physically irrelevant. The cohomology of }$%
\mathcal{B}_\Sigma \;$\textit{is called empty if all solutions of eq.}$%
\left( \ref{def-coh}\right) \;$\textit{are of the type }$\left( \ref
{triv-coh}\right) $.

\vspace{5mm}

Acting now on both sides of eq.$\left( \ref{broken-q-s-t}\right) \;$with the
operator $\mathcal{B}_\Gamma \;$

\begin{equation}
\mathcal{B}_\Gamma =\int d^4x\left( \frac{\delta \Gamma }{\delta \varphi ^i}%
\frac \delta {\delta \varphi ^{*i}}\;+\frac{\delta \Gamma }{\delta \varphi
^{*i}}\frac \delta {\delta \varphi ^i}\right) \;+\;\frac
12C_{\;AB}^C\varepsilon ^A\varepsilon ^B\frac \partial {\partial \varepsilon
^C}\;=\mathcal{B}_\Sigma +\;O(\hbar )\;,  \label{b-gamma-op}
\end{equation}
and making use of

\begin{equation}
\mathcal{B}_\Gamma \mathcal{S(}\Gamma \mathcal{)=}\;0\;,  \label{q-ex-rel}
\end{equation}
we get, to the lowest order in $\hbar $, the consistency condition

\begin{equation}
\mathcal{B}_\Sigma \Delta =0\,\;.  \label{an-cons-cond}
\end{equation}
The latter is nothing but a cohomology problem for the operator $\mathcal{B}%
_\Sigma \;$in the sector of the integrated local field polynomials with
ghost number one. Let us now suppose that the most general solution of the
consistency condition $\left( \ref{an-cons-cond}\right) $\ can be written in
the exact form

\begin{equation}
\Delta =\mathcal{B}_\Sigma \widehat{\Delta }\;,  \label{triv-sol}
\end{equation}
for some local integrated polynomial $\widehat{\Delta }$\ with ghost number
zero. Therefore, the redefined vertex functional

\begin{equation}
\overline{\Gamma }=\Gamma -\hbar ^n\widehat{\Delta }\;,
\label{red-vert-funct}
\end{equation}
obeys the Slavnov-Taylor identity

\begin{equation}
\mathcal{S(}\overline{\Gamma }\mathcal{)=}O(\hbar ^{n+1})\;.
\label{improv-s-t}
\end{equation}
This equation means that if the breaking term $\Delta \;$is cohomologically
trivial, one can always extend the Slavnov-Taylor identity to the order $%
\hbar ^n$. The procedure can be iterated, allowing us to conclude that if
the cohomology of $\mathcal{B}_\Sigma \;$is empty in the sector of ghost
number one it is always possible to implement at the quantum level the
classical Slavnov-Taylor identity $\left( \ref{class-s-t}\right) $.\ In this
case the model is said to be anomaly free. On the contrary, when the
cohomology of $\mathcal{B}_\Sigma \;$is not empty, \textit{i.e.}

\begin{eqnarray}
\Delta &=&r\mathcal{A}+\mathcal{B}_\Sigma \widehat{\Delta }\;\;,\;
\label{nonv-coh} \\
\mathcal{A} &\neq &\;\mathcal{B}_\Sigma \;\widehat{\mathcal{A}}\;,  \nonumber
\end{eqnarray}
with $r$ an arbitrary coefficient and $\widehat{\mathcal{A}}\;$some local
field polynomial, it is not possible to compensate the breaking term by
adding suitable local terms to the vertex functional. The best that one can
do is just to reabsorb the trivial part $\mathcal{B}_\Sigma \widehat{\Delta }%
\;$of eq.$\left( \ref{nonv-coh}\right) $,

\begin{equation}
\mathcal{S(}\Gamma )\;\mathcal{=}r\hbar ^n\mathcal{A}+O(\hbar ^{n+1})\;.
\label{an-s-t}
\end{equation}
In this case one speaks of an anomaly, meaning that the classical symmetries
cannot be implemented at the quantum level.

\begin{remark}
\textit{It is important here to underline that the algebraic method does not
provide the numerical value of the coefficient }$r$. \textit{This means that
the anomaly }$\mathcal{A}$\textit{\ appearing in the left hand side of eq.}$%
\left( \ref{an-s-t}\right) $\textit{\ is only a potential anomaly, whose
existence has to be confirmed with an explicit computation of }$r$. \textit{%
Moreover, the vanishing of the coefficient }$r$ \textit{does not imply the
absence of the anomaly. It only means that the anomaly is absent at the
order }$\hbar ^n$.\textit{\ Possible anomalous contributions are expected at
higher orders, unless one is able to establish a nonrenormalization theorem.
This is the case, for instance, of the Adler-Bardeen nonrenormalization
theorem of the gauge anomaly which states that if the coefficient }$r$%
\textit{\ is vanishing at the one loop order, it will vanish at all orders 
\cite{adba}. }
\end{remark}

Having discussed the characterization of the possible anomalous terms, let
us turn now to the analysis of the invariant counterterms, \textit{i.e.} of
the local ambiguities which affect the Slavnov-Taylor identity. In fact, if $%
\Gamma \;$is a vertex functional which satisfies the Slavnov-Taylor identity
to the order $\hbar ^n$

\begin{equation}
\mathcal{S(}\Gamma \mathcal{)}=O(\hbar ^{n+1})\;,  \label{s-t-amb}
\end{equation}
then adding to $\Gamma $ any local invariant field polynomial $\hbar
^n\Theta $ with the same quantum numbers and dimension of the classical
action $\Sigma \;$

\begin{equation}
\mathcal{B}_\Sigma \Theta =0\;,  \label{count}
\end{equation}
the resulting vertex functional still satisfies eq.$\left( \ref{s-t-amb}%
\right) $, \textit{i.e.}

\begin{eqnarray}
\Gamma _\Theta &=&\Gamma +\hbar ^n\Theta \;,  \label{g-th} \\
\mathcal{S(}\Gamma _\Theta \mathcal{)} &=&O(\hbar ^{n+1})\;.  \nonumber
\end{eqnarray}
In other words, the Slavnov-Taylor identity characterizes the vertex
functional $\Gamma \;$only up to local invariant polynomials $\Theta $ which
can be freely added to each order of perturbation theory. The vertex
functional $\Gamma $ will be uniquely fixed only once the most general
solution of eq.$\left( \ref{count}\right) $ has been given and a suitable
set of renormalization conditions has been imposed. Again, eq.$\left( \ref
{count}\right) $ shows that the search of the invariant counterterms is a
problem of cohomology of $\mathcal{B}_\Sigma $ in the space of the
integrated local field polynomials with ghost number zero. In general, $%
\Theta \;$will be of the form

\begin{equation}
\Theta =\Theta ^{coh}+\mathcal{B}_\Sigma \widehat{\Theta }\;,
\label{count-coh}
\end{equation}
with $\Theta ^{coh}\;$identifying the nontrivial cohomology sectors of $%
\mathcal{B}_\Sigma $.

Let us now introduce the notion of \textit{stability} of the classical
action. The complete action $\Sigma \;$of eq.$\left( \ref{comp-act}\right) $
is said to be \textit{stable} if the most general local invariant
counterterm can be reabsorbed by means of a redefinition of the fields and
of the parameters, \textit{i.e.} denoting with $\left\{ g\right\} $ the
parameters of $\Sigma $ (coupling constants, masses, gauge parameters,
etc....) we have

\begin{equation}
\Sigma (\varphi ^i,\varphi ^{*i},\varepsilon ^A,g)+\hbar ^n\Theta =\Sigma
(\varphi _0^i,\varphi _0^{*i},\varepsilon _0^A,g_0)+O(\hbar ^{n+1})\;,
\label{stab}
\end{equation}
with

\begin{eqnarray}
\varphi _0^i &=&\varphi ^i(1+\hbar ^n\mathcal{\varsigma }_\varphi )+O(\hbar
^{n+1})\;,  \label{ren-const} \\
\varphi _0^{*i} &=&\varphi ^{*i}(1+\hbar ^n\mathcal{\varsigma }_{\varphi
^{*}})+O(\hbar ^{n+1})\;,  \nonumber \\
\varepsilon _0^A &=&\varepsilon ^A(1+\hbar ^n\mathcal{\varsigma }%
_\varepsilon )+O(\hbar ^{n+1})\;,  \nonumber \\
g_0 &=&g(1+\hbar ^n\mathcal{\varsigma }_g)+O(\hbar ^{n+1})\;,  \nonumber
\end{eqnarray}
$\left( \mathcal{\varsigma }_\varphi \mathcal{\varsigma }_{\varphi ^{*}}%
\mathcal{\varsigma }_\varepsilon \mathcal{\varsigma }_g\right) $\ being 
\textit{renormalization} \textit{constants}. Let us also remark that the
knowledge of the nontrivial counterterms $\Theta ^{coh}\;$has a very
important meaning. One can show indeed that the elements of $\Theta ^{coh}\;$%
correspond to the renormalization of the physical parameters of $\Sigma $, 
\textit{i.e.} of the coupling constants and of the masses, while the trivial
term $\mathcal{B}_\Sigma \widehat{\Theta }\;$turns out to be related to the
unphysical renormalization of the field amplitudes and of the gauge
parameters. In addition, the renormalized Green's functions with the
insertion of local gauge invariant composite operators can be proven to be
independent from the parameters belonging to the trivial part $\mathcal{B}%
_\Sigma \widehat{\Theta }$.

\begin{definition}
\textit{The classical action }$\Sigma \;$\textit{satisfying the classical
Slavnov-Taylor identity }$\left( \ref{class-s-t}\right) $ \textit{is said to
be renormalizable if the following two items are fulfilled, namely\\ i)
there are no anomalies, i.e.}
\end{definition}

\begin{eqnarray}
\Sigma &\rightarrow &\Gamma =\Sigma +O(\hbar )\;,  \label{abs-anom} \\
\mathcal{S(}\Gamma \mathcal{)} &=&0\;,  \nonumber
\end{eqnarray}
\textit{ii) the action is stable.\\The absence of anomalies ensures that the
classical symmetries can be implemented at the quantum level, while the
stability means that all possible local countertems compatible with the
symmetry content can be reabsorbed by redefining the fields and the
parameters of the original action }$\Sigma $.

\textit{\vspace{3mm}}

In summary, we have seen that the renormalizability of a given classical
model can be established by looking at the cohomology of the operator $%
\mathcal{B}_\Sigma \;$in the sector of the integrated local field
polynomials with ghost number respectively one (anomalies) and zero
(counterterms).

\begin{remark}
\textit{We should also mention that in the case in which one (or more) of
the global generators }$\delta _A$ \textit{acts linearly on the quantum
fields }$\varphi ^i$\textit{, the dependence of the quantum action }$\Gamma
\;$\textit{from the corresponding global ghost turns out to be uniquely
fixed already at the classical level. This means that, denoting with }$%
\delta _C^l\;$\textit{the linearly realized global generator, for the
corresponding global ghost }$\varepsilon ^{lC}$ \textit{we may write the
following classical identity } 
\begin{equation}
\frac{\partial \Sigma }{\partial \varepsilon ^{lC}}=\Delta _C^l\;=\int d^4x%
\mathcal{M}_C^{li}\varphi ^i\;,  \label{lin-br-w}
\end{equation}
\textit{where }$\mathcal{M}_C^{li}\;$\textit{denote a set of generalized
coefficients depending only on the antifields }$\varphi ^{*i}$\textit{,\ on
the global ghosts }$\varepsilon ^A$ \textit{and on their space-time
derivatives. The breaking }$\Delta _C^l$\textit{, being linear in the
quantum fields }$\varphi ^i$\textit{, is thus a classical breaking and will
be not affected by the quantum corrections \cite{book}}.\textit{\ Therefore
the equation }$\left( \ref{lin-br-w}\right) $ \textit{has the meaning} 
\textit{of a linearly broken Ward identity which, once extended at the
quantum level, will imply that the higher order terms of the renormalized
vertex functional }$\Gamma $%
\begin{equation}
\Gamma =\Sigma +\sum_{j=1}^\infty \hbar ^j\Gamma ^j\;,  \label{q-e-a}
\end{equation}
\textit{do not depend from the global ghost }$\varepsilon ^{lC}$.\textit{\
In fact, from } 
\begin{equation}
\frac{\partial \Gamma }{\partial \varepsilon ^{lC}}=\Delta _C^l\;,
\label{q-ext}
\end{equation}
\textit{it follows that } 
\begin{equation}
\frac{\partial \Gamma ^j}{\partial \varepsilon ^{lC}}=0\;,\;\;\;\;\;\;\;j%
\geq 1\;,  \label{q-ind}
\end{equation}
\textit{due to equation }$\left( \ref{lin-br-w}\right) .$ \textit{This
result shows that the dependence of the theory from the global ghosts
corresponding to linearly realized symmetries is completely fixed by the
classical equation }$\left( \ref{lin-br-w}\right) .$ \textit{As we shall see
later on in the analysis of TYM, this will be the case of the global ghost
associated to the space-time translation invariance. }
\end{remark}

\subsection{Some useful result on cohomology}

Let us conclude this short introduction to the \textit{Algebraic
Renormalization} by stating some useful result on the computation of the
cohomology of the operator $\mathcal{B}_\Sigma $. Let us begin by
underlining the important role played by the functional space the operator $%
\mathcal{B}_\Sigma $ acts upon. Different functional spaces yield, in
general, different cohomology classes for $\mathcal{B}_\Sigma $. In the
previous Subsection we have adopted as basic functional space for the
operator $\mathcal{B}_\Sigma $ the space of the integrated local polynomials
in the fields $\varphi $, antifields $\varphi ^{*}$, global ghosts $%
\varepsilon $ and their space-time derivatives. We emphasize here that the
choice of this functional space follows directly from the locality
properties of the renormalized perturbation theory.

\begin{remark}
\textit{Concerning in particular the global ghosts }$\varepsilon ^A$\textit{%
\ it will be very easy to check that the Feynman rules stemming from the
quantized TYM action yield a perturbative expansion which is analytic in the 
}$\varepsilon ^A$\textit{'s. This analyticity property, whose precise
mathematical meaning is that of a formal power series, will be of great
importance in order to understand the BRST cohomology classes of TYM. }
\end{remark}

On the space of the integrated local field polynomials the operator $%
\mathcal{B}_\Sigma \;$has a natural decomposition as

\begin{eqnarray}
\mathcal{B}_\Sigma &=&b_0+b_R\;,  \label{dec} \\
b_0^2 &=&0\;,  \nonumber
\end{eqnarray}
$b_0$ being the so called abelian approximation.

\begin{example}
\textit{Let }$s$\textit{\ be the nilpotent operator of the eq.}$\left( \ref
{brs-tr}\right) $\textit{\ acting on the space of the local polynomials in
the variables }$\left( A_\mu ,\left\{ \lambda \right\} ,c\right) $.\ \textit{%
Therefore} 
\begin{equation}
s=s_0+s_R\;,  \label{brs-dec}
\end{equation}
\textit{with } 
\begin{equation}
s_0A_\mu =-\partial _\mu c\;,\;\;\;\;\;\;\;s_0c=0\;,\;\;\;\;\;\;\;s_0\lambda
=0\;,\;\;\;\;\;\;s_0^2=0\;,\;\;  \label{s0}
\end{equation}
\textit{and } 
\begin{equation}
s_RA_\mu =[c,A_\mu ]\;,\;\;\;\;\;\;s_Rc=c^2\;,\;\;\;\;\;\;s_R\lambda
=[c,\lambda ]\;.  \label{sR}
\end{equation}
\textit{One sees thus that }$s_0$ \textit{corresponds in fact to the abelian
approximation in which all the commutators }$[\cdot ,\cdot ]$ \textit{have
been ignored. }
\end{example}

The usefulness of the decomposition $\left( \ref{dec}\right) $ relies on a
very general theorem stating that the cohomology of the complete operator $%
\mathcal{B}_\Sigma $ is isomorphic to a subspace of the cohomology of the
operator $b_0$. In most cases this result allows to obtain a large amount of
informations on the cohomology of $\mathcal{B}_\Sigma $ by analysing that of
the simpler operator $b_0$. Let us also remark that the aforementioned
theorem, although referred here to the abelian approximation, is valid for
other kinds of decomposition of the operator $\mathcal{B}_\Sigma $.

Let us give now a second important result, known as the doublets theorem. A
pair of fields $\left( u,v\right) $ is called a BRST\ doublet if

\begin{equation}
b_0u=v\;,\;\;\;\;\;\;b_0v=0\;.  \label{doublet}
\end{equation}
It can be shown that if two fields appear in a BRST doublet, then the
cohomology of $b_0$, and therefore that of $\mathcal{B}_\Sigma $, does not
depend on these fields. This second result allows to eliminate from the game
all the fields appearing as BRST\ doublets, greatly simplifying the
computation of the cohomology classes.

More details on cohomology can be found in Chapter 5 of ref. \textit{\cite
{book}}.

\newpage\ 

\section{Witten's topological Yang-Mills theory}

\subsection{The action and its fermionic symmetry}

Topological Yang-Mills theory has been proposed at the end of the eighty's
with the aim of providing a field theory framework for the topological
invariants of euclidean four manifolds \textit{\cite{tym}}. The model allows
in fact to define a set of observables, \textit{i.e. }local field
polynomials integrated over suitable homology cycles, whose correlation
functions turn out to be deeply related with the so called Donaldson
invariants \textit{\cite{donald}}.

Although TYM can be formulated on smooth four manifolds, we shall consider
here the case of the flat euclidean space-time\footnote{%
See \textit{\cite{ans}} for the computation of some topological invariant
associated to submanifolds of $R^4$.}. In fact our attitude in these
lectures is to interpret TYM theory as a twisted version of the conventional 
$N=2$ supersymmetric euclidean Yang-Mills theory, as it will become clear in
the next Sections where the relationship with the cohomological formulations
of Labastida-Pernici \textit{\cite{lp} }and Baulieu-Singer \textit{\cite{bs}}
will be\textit{\ }also discussed\textit{. }Following the original Witten's
work, the TYM classical action is given by

\begin{eqnarray}
\mathcal{S}_{TYM} &=&\frac 1{g^2}tr\displaystyle\int d^4x\;\left( \frac
12F_{\mu \nu }^{+}F^{+\mu \nu }\;-\chi ^{\mu \nu }(D_\mu \psi _\nu -D_\nu
\psi _\mu )^{+}\;\right.  \label{tym} \\
&&\;\;\;\;\;\;\;\;\;\;\;\;\;\;\;\;\;+\eta D_\mu \psi ^\mu \;-\frac 12%
\overline{\phi }D_\mu D^\mu \phi \;+\frac 12\overline{\phi }\left\{ \psi
^\mu ,\psi _\mu \right\}  \nonumber \\
&&\;\;\;\;\;\;\;\;\;\;\;\;\left. \;\;\;-\frac 12\phi \left\{ \chi ^{\mu \nu
},\chi _{\mu \nu }\right\} \;-\frac 18\left[ \phi ,\eta \right] \eta -\frac
1{32}\left[ \phi ,\overline{\phi }\right] \left[ \phi ,\overline{\phi }%
\right] \right) \;,  \nonumber
\end{eqnarray}
where $g$ is the \textit{unique} coupling constant and $F_{\mu \nu }^{+}$ is
the self-dual part of the Yang-Mills field strength

\begin{eqnarray}
F_{\mu \nu }^{+} &=&F_{\mu \nu }+\frac 12\varepsilon _{\mu \nu \rho \sigma
}F^{\rho \sigma }\;,\;\;\;\;\;\;\widetilde{F}_{\mu \nu }^{+}=\frac
12\varepsilon _{\mu \nu \rho \sigma }F^{+\rho \sigma }=F_{\mu \nu }^{+}\;,
\label{f+} \\
F_{\mu \nu } &=&\partial _\mu A_\nu -\partial _\nu A_\mu +\left[ A_\mu
,A_\nu \right] \;,  \nonumber
\end{eqnarray}
$\varepsilon _{\mu \nu \rho \sigma }$ being the totally antisymmetric
Levi-Civita tensor

\begin{equation}
\varepsilon _{\mu \nu \rho \sigma }\varepsilon ^{\rho \sigma \tau \lambda
}=2(\delta _\mu ^\tau \delta _\nu ^\lambda -\delta _\nu ^\tau \delta _\mu
^\lambda )\;.  \label{eps-t}
\end{equation}
The three fields \ $\left( \chi _{\mu \nu },\psi _\mu ,\eta \right) $ in the
expression $\left( \ref{tym}\right) $ are anticommuting with $\chi _{\mu \nu
}$ self-dual

\begin{equation}
\widetilde{\chi }_{\mu \nu }=\frac 12\varepsilon _{\mu \nu \rho \sigma }\chi
^{\rho \sigma }=\chi _{\mu \nu }=-\chi _{\nu \mu }\;.  \label{chi}
\end{equation}
Accordingly, the term $(D_\mu \psi _\nu -D_\nu \psi _\mu )^{+}\;$in $\left( 
\ref{tym}\right) $ has to be understood as

\begin{equation}
(D_\mu \psi _\nu -D_\nu \psi _\mu )^{+}=(D_\mu \psi _\nu -D_\nu \psi _\mu
)+\frac 12\varepsilon _{\mu \nu \rho \sigma }(D^\rho \psi ^\sigma -D^\sigma
\psi ^\rho )\;,  \label{Dpsi}
\end{equation}
$\left( D_\mu \cdot =\partial _\mu \cdot +[A_\mu ,\cdot ]\right) $ being the
covariant gauge derivative. Finally $\left( \phi ,\overline{\phi }\right) $
are commuting complex scalar fields, $\overline{\phi }$ being assumed to be
the complex conjugate\footnote{%
See also Witten's remark on this point given at the end of the Subsect. 2.2
of \textit{\cite{tym}}.} of $\phi $.

Of course, TYM being a gauge theory, is left invariant by the gauge
transformations

\begin{eqnarray}
\delta _\epsilon ^gA_\mu &=&-D_\mu \epsilon \;,  \label{g-transf} \\
\delta _\epsilon ^g\lambda &=&\left[ \epsilon ,\lambda \right]
\;,\;\;\;\;\lambda =\chi ,\psi ,\eta ,\phi ,\overline{\phi }\;,  \nonumber
\end{eqnarray}

\begin{equation}
\delta _\epsilon ^g\mathcal{S}_{TYM}=0\;,  \label{g-inv}
\end{equation}

\begin{remark}
\textit{It is easily checked that the kinetic terms in the action }$\left( 
\ref{tym}\right) $\textit{\ corresponding to the fields }$\left( \chi ,\psi
,\eta ,\phi ,\overline{\phi }\right) $ \textit{are nondegenerate, so that
these fields have well defined propagators. The only degeneracy is that
related to the pure Yang-Mills term }$F_{\mu \nu }^{+}F^{+\mu \nu }$. 
\textit{Therefore}, \textit{from eq.}$\left( \ref{g-transf}\right) $ \textit{%
one is led to interpret the fields }$\left( \chi ,\psi ,\eta ,\phi ,%
\overline{\phi }\right) $ \textit{as ordinary matter fields, in spite of the
unconventional tensorial character of }$\left( \chi _{\mu \nu },\psi _\mu
\right) $\textit{. This point will become more clear later on, once the
relationship between TYM and the }$N=2\;$\textit{euclidean gauge theories
will be established. }
\end{remark}

In addition to the gauge invariance, the action $\left( \ref{tym}\right) $
turns out to be left invariant by the following nonlinear transformations 
\textit{\cite{tym}}

\begin{eqnarray}
\delta _{\mathcal{W}}A_\mu &=&\psi _\mu \;,  \label{d-transf} \\
\delta _{\mathcal{W}}\psi _\mu &=&-D_\mu \phi \;,  \nonumber \\
\delta _{\mathcal{W}}\phi &=&0\;,  \nonumber \\
\delta _{\mathcal{W}}\chi _{\mu \nu } &=&F_{\mu \nu }^{+}\;,  \nonumber \\
\delta _{\mathcal{W}}\overline{\phi } &=&2\eta \;,  \nonumber \\
\delta _{\mathcal{W}}\eta &=&\frac 12\left[ \phi ,\overline{\phi }\right] \;,
\nonumber
\end{eqnarray}
and

\begin{equation}
\delta _{\mathcal{W}}\mathcal{S}_{TYM}=0\;.  \label{d-inv}
\end{equation}
The operator $\delta _{\mathcal{W}}$ is of fermionic type and obeys the
relation

\begin{equation}
\delta _{\mathcal{W}}^2=\delta _\phi ^g\;+\;(\chi \text{\textrm{-eq. of
motion}})\;,  \label{d2}
\end{equation}
$\delta _\phi ^g\;$denoting a gauge transformation with gauge parameter $%
\phi .$ More precisely

\begin{eqnarray}
\delta _{\mathcal{W}}^2A_\mu &=&-D_\mu \phi \;,  \label{d2-expl} \\
\delta _{\mathcal{W}}^2\psi _\mu &=&\left[ \phi ,\psi _\mu \right] \;, 
\nonumber \\
\delta _{\mathcal{W}}^2\phi &=&0\;,  \nonumber \\
\delta _{\mathcal{W}}^2\overline{\phi } &=&\left[ \phi ,\overline{\phi }%
\right] \;,  \nonumber \\
\delta _{\mathcal{W}}^2\eta &=&\left[ \phi ,\eta \right] \;,  \nonumber \\
\delta _{\mathcal{W}}^2\chi _{\mu \nu } &=&\left[ \phi ,\chi _{\mu \nu
}\right] -\frac{g^2}2\frac{\delta \mathcal{S}_{TYM}}{\delta \chi ^{\mu \nu }}%
\;.  \nonumber
\end{eqnarray}
The equations $\left( \ref{d2}\right) ,\left( \ref{d2-expl}\right) $\ mean
essentially that the operator $\delta _{\mathcal{W}}$ becomes nilpotent when
acting on the space of the gauge invariant functionals. This property turns
out to play an important role in the construction of the Witten's
observables. In particular, eq.$\left( \ref{d2}\right) $ shows that the
operator $\delta _{\mathcal{W}}\;$belongs to the class of the operators $%
\left\{ \delta _A\right\} \;$considered in the previous Section (see eq.$%
\left( \ref{add-algebra}\right) $).

\begin{remark}
\textit{One should remark that the relative coefficients of the various
terms of the TYM action }$\left( \ref{tym}\right) $ \textit{are not
completely fixed by the fermionic symmetry }$\delta _{\mathcal{W}}$\textit{.
In other words, the action }$\left( \ref{tym}\right) $ \textit{is not the
most general gauge invariant action compatible with the }$\delta _{\mathcal{W%
}}$-\textit{invariance. Nevertheless, we shall see that }$\mathcal{S}_{TYM}$ 
\textit{turns out to possess additional nonlinear invariances which fix
completely the relative numerical coefficients of the various terms of
expression }$\left( \ref{tym}\right) $ \textit{and allow for a unique
coupling constant. Moreover, these additional nonlinear symmetries give rise
together with the }$\delta _{\mathcal{W}}$-\textit{symmetry to a twisted
version of the} $N=2$ \textit{susy algebra in the Wess-Zumino gauge. }
\end{remark}

Following Witten, it is also easily seen that assigning to $\left( A,\chi
,\psi ,\eta ,\phi ,\overline{\phi }\right) $ the following $\mathcal{R}$%
-charges $\left( 0,-1,1,-1,2,-2\right) ,\;$the TYM action $\left( \ref{tym}%
\right) $ has vanishing total $\mathcal{R}$-charge.

Let us display, finally, the quantum numbers of all the fields and of $%
\delta _{\mathcal{W}}$.

\[
\stackunder{Table\;1.}{\stackrel{Dim.and\text{ }\mathcal{R}-charges}{
\begin{tabular}{|c|c|c|c|c|c|c|c|}
\hline
& $A_\mu $ & $\chi _{\mu \nu }$ & $\psi _\mu $ & $\eta $ & $\phi $ & $%
\overline{\phi }$ & $\delta _{\mathcal{W}}$ \\ \hline
$\mathrm{dim}.$ & $1$ & $3/2$ & $3/2$ & $3/2$ & $1$ & $1$ & $1/2$ \\ \hline
$\mathcal{R}-\mathrm{charg}.$ & $0$ & $-1$ & $1$ & $-1$ & $2$ & $-2$ & $1$
\\ \hline
$\mathrm{nature}$ & $comm$. & $ant.$ & $ant.$ & $ant.$ & $comm.$ & $comm.$ & 
$ant.$ \\ \hline
\end{tabular}
}} 
\]

\subsection{Twisting the N=2 supersymmetric algebra}

For a better understanding of the TYM action $\left( \ref{tym}\right) $ and
of its fermionic symmetry $\left( \ref{d-inv}\right) $, let us present now
the twisting procedure of the $N=2$ supersymmetric algebra in flat euclidean
space-time. We shall follow here mainly the detailed analysis done by M.
Mari\~{n}o\footnote{%
This work is an uptodate reference on this subject, including a discussion
of the twisting procedure in the presence of matter multiplets and central
charges, as well as a study of the relationship between topological field
theories and gauge models with extended supersymmetry.} \textit{\cite{mar}}.
In the absence of central extension\footnote{%
This is the case for instance of theories involving only massless fields.}
the $N=2$ supersymmetry is characterized by $8$ charges $\left( \mathcal{Q}%
_{\;\alpha }^i,\overline{\mathcal{Q}}_{\;\dot{\alpha}}^j\right) $ obeying
the following relations

\begin{eqnarray}
\left\{ \mathcal{Q}_{\;\alpha }^i,\overline{\mathcal{Q}}_{j\dot{\alpha}%
}\right\} &=&\delta _j^i\left( \sigma ^\mu \right) _{\alpha \dot{\alpha}%
}\partial _{\mu \;,}  \label{n=2} \\
\left\{ \mathcal{Q}_{\;\alpha }^i,\mathcal{Q}_{\;\beta }^j\right\}
&=&\left\{ \overline{\mathcal{Q}}_{\;\dot{\alpha}}^i,\overline{\mathcal{Q}}%
_{\;\dot{\beta}}^j\right\} =0\;,  \nonumber
\end{eqnarray}
where $(\alpha ,\dot{\alpha})=1,2$ are the spinor indices, $(i,j)=1,2\;$the
internal $SU(2)$ indices labelling the different charges of $N=2,$ and $%
\left( \sigma ^\mu \right) _{\alpha \dot{\alpha}}=(1,\stackrel{\rightarrow }{%
i\sigma }),\;\stackrel{\rightarrow }{\sigma }\;$being the Pauli matrices\
(see App.A for the euclidean susy conventions). The special feature of $N=2$
is that both spinor and internal indices run from $1\;$to $2$, making
possible to identify the index $i$ with one of the two spinor indices $%
(\alpha ,\dot{\alpha})$. It is precisely this identification which defines
the twisting procedure. As explained in \textit{\cite{mar}}, this is
equivalent to a redefinition of the action of the four dimensional rotation
group. Indeed, in the four dimensional flat euclidean space-time the global
symmetry group of $N=2$ supersymmetry is $SU(2)_L\times SU(2)_R\times
SU(2)_I\times U(1)_{\mathcal{R}}\;$where $SU(2)_L\times SU(2)_R$\ is the
rotation group, and $SU(2)_I\;$and $U(1)_{\mathcal{R}}\;$are the symmetry
groups corresponding respectively to $SU(2)$-transformations of the internal
index $i$ and to the $\mathcal{R}$-symmetry, the $\mathcal{R}$-charges of $%
\left( \mathcal{Q}_{\;\alpha }^i,\overline{\mathcal{Q}}_{\;\dot{\alpha}%
}^j\right) \;$being respectively $(1,-1)$. The twisting procedure consists
thus of replacing the rotation group $SU(2)_L\times SU(2)_R\;$by $%
SU(2)_L\times SU(2)_R^{\prime }\;$where $SU(2)_R^{\prime }$ is the diagonal
sum of $SU(2)_R\;$and of $SU(2)_I$. Identifying therefore the internal index 
$i\;$with the spinor index $\alpha $, the $N=2\;$susy algebra $\left( \ref
{n=2}\right) \;$becomes

\begin{eqnarray}
\left\{ \mathcal{Q}_{\;\alpha }^\beta ,\overline{\mathcal{Q}}_{\gamma \dot{%
\alpha}}\right\} &=&\delta _\gamma ^\beta \left( \sigma ^\mu \right)
_{\alpha \dot{\alpha}}\partial _{\mu \;,}  \label{tn=2} \\
\left\{ \mathcal{Q}_{\;\alpha }^\beta ,\mathcal{Q}_{\;\gamma }^\delta
\right\} &=&\left\{ \overline{\mathcal{Q}}_{\alpha \dot{\alpha}},\overline{%
\mathcal{Q}}_{\beta \dot{\beta}}\right\} =0\;.  \nonumber
\end{eqnarray}
Let us define now the following generators\ $\left( \delta ,\delta _\mu
,\delta _{\mu \nu }\right) $ with $\mathcal{R}$-charge $(1,-1,1)$
respectively;

\begin{eqnarray}
\delta &=&\frac 1{\sqrt{2}}\varepsilon ^{\alpha \beta }\mathcal{Q}_{\beta
\alpha }\;,  \label{t-charges} \\
\delta _\mu &=&\frac 1{\sqrt{2}}\overline{\mathcal{Q}}_{\alpha \dot{\alpha}%
}\;\left( \overline{\sigma }_\mu \right) ^{\dot{\alpha}\alpha }\;,  \nonumber
\\
\delta _{\mu \nu } &=&\frac 1{\sqrt{2}}(\sigma _{\mu \nu })^{\alpha \beta }%
\mathcal{Q}_{\beta \alpha }=-\delta _{\nu \mu \;},  \nonumber
\end{eqnarray}
where, as usual,

\begin{equation}
\sigma _{\mu \nu }=\frac 12(\sigma _\mu \overline{\sigma }_\nu -\sigma _\nu 
\overline{\sigma }_\mu )\;.  \label{smn}
\end{equation}
Notice that the generators $\delta _{\mu \nu }$ are self-dual

\begin{equation}
\delta _{\mu \nu }=\widetilde{\delta }_{\mu \nu }=\frac 12\varepsilon _{\mu
\nu \rho \sigma }\delta ^{\rho \sigma }\;,  \label{self-d}
\end{equation}
due to the fact that the matrices $\sigma _{\mu \nu }$ are self-dual in
euclidean space-time (see App.A). Equations $\left( \ref{t-charges}\right) $
show that we can replace the spinorial charges $\left( \mathcal{Q}_{\;\alpha
}^i,\overline{\mathcal{Q}}_{\;\dot{\alpha}}^j\right) $ with the $8$
generators $\left( \delta ,\delta _\mu ,\delta _{\mu \nu }\right) $,
respectively a scalar $\delta $, a vector $\delta _\mu \;$and a self-dual
tensor $\delta _{\mu \nu }$. In terms of these generators, the $N=2$ susy
algebra reads now

\begin{eqnarray}
\delta ^2 &=&0\;,  \label{talg-1} \\
\left\{ \delta ,\delta _\mu \right\} &=&\partial _\mu \;,  \nonumber \\
\left\{ \delta _\mu ,\delta _\nu \right\} &=&0\;,  \nonumber
\end{eqnarray}
and

\begin{eqnarray}
\left\{ \delta _\mu ,\delta _{\rho \sigma }\right\} &=&-(\varepsilon _{\mu
\rho \sigma \nu }\partial ^\nu +g_{\mu \rho }\partial _\sigma -g_{\mu \sigma
}\partial _\rho )\;,  \label{talg-2} \\
\left\{ \delta ,\delta _{\mu \nu }\right\} &=&\left\{ \delta _{\mu \nu
},\delta _{\rho \sigma }\right\} =0\;,  \nonumber
\end{eqnarray}
where $g_{\mu \nu }=diag(+,+,+,+)$ is the flat euclidean metric and where
use has been made of the relations (see App.A)

\begin{eqnarray}
tr(\sigma ^\mu \overline{\sigma }^\nu ) &=&2g^{\mu \nu },  \label{s-rel} \\
\sigma _{\mu \nu }\sigma _\rho &=&-(\varepsilon _{\mu \nu \rho \tau }\sigma
^\tau +g_{\mu \rho }\sigma _\nu -g_{\nu \rho }\sigma _\mu )\;.  \nonumber
\end{eqnarray}
The charges $\left( \delta ,\delta _\mu ,\delta _{\mu \nu }\right) \;$are
called twisted generators and the eqs.$\left( \ref{talg-1}\right) $,$%
\;\left( \ref{talg-2}\right) $ define the twisted version of the $N=2\;$susy
algebra $\left( \ref{n=2}\right) $. We see, in particular, that the singlet
operator $\delta $, although rather different from the operator $s$ of the
gauge transformations $\left( \ref{brs-tr}\right) $, is nilpotent, allowing
thus for a BRST-charge like interpretation. It should also be remarked from
eq.$\left( \ref{talg-1}\right) $ that the singlet and the vector generators $%
(\delta ,\delta _\mu )$ give rise to an algebraic structure which is typical
of the topological models \textit{\cite{tft}}. In this case the vector
charge $\delta _\mu $, usually called vector supersymmetry, is known to play
an important role in the derivation of the ultraviolet finiteness properties
of the topological models and in the construction of their observables 
\textit{\cite{tft}}. Therefore the $N=2$ susy algebra, when rewritten in
terms of the twisted generators, displays the same algebraic structure of
the topological models. In what follows we shall check that this feature is
more than a simple analogy. Let us emphasize in fact that, since a few
years, the relationship between topological field theories and models with
extended supesrymmetry is becoming more and more apparent \textit{\cite{mar}}%
.

\begin{remark}
It is useful to remark here that the \ supersymmetric charges $\left( 
\mathcal{Q}_{\;\alpha }^i,\overline{\mathcal{Q}}_{\;\dot{\alpha}}^j\right) $%
\ of eq.$\left( \ref{n=2}\right) $\ have been implicitly referred to a
linear realization of supersymmetry, meaning that they act linearly on the
components of the $N=2$\ multiplets. Instead, in what follows we shall deal
with a different situation in which supersymmetry in nonlinearly realized,
due to the use of the Wess-Zumino gauge \textit{\cite{white,magg}}. As it is
well known, the Wess-Zumino gauge allows to reduce the number of fields
component, simplifying considerably the full analysis. There is however a
price to pay. The supersymmetric transformations are now nonlinear and the
algebra between the supersymmetric charges closes on the translations only
modulo gauge transformations and equations of motion. Accordingly, the $N=2$%
\ algebra in the Wess-Zumino gauge will read then 
\begin{eqnarray}
\left\{ \mathcal{Q}_{\;\alpha }^i,\overline{\mathcal{Q}}_{j\dot{\alpha}%
}\right\}  &=&\delta _j^i\left( \sigma ^\mu \right) _{\alpha \dot{\alpha}%
}\partial _\mu +(\mathrm{gauge}\;\mathrm{transf.)\;}+\;(\mathrm{eqs.}\;%
\mathrm{of\;mot.})\;,  \label{wz-n=2} \\
\left\{ \mathcal{Q}_{\;\alpha }^i,\mathcal{Q}_{\;\beta }^j\right\} 
&=&\left\{ \overline{\mathcal{Q}}_{\dot{\alpha}}^i,\overline{\mathcal{Q}}_{%
\dot{\beta}}^j\right\} =(\mathrm{gauge}\;\mathrm{transf.)\;}+\;(\mathrm{eqs.}%
\;\mathrm{of\;mot.})\;.  \nonumber
\end{eqnarray}
Analogously, for the twisted version\ 
\begin{eqnarray}
\delta ^2 &=&(\mathrm{gauge}\;\mathrm{transf.)\;}+\;(\mathrm{eqs.}\;\mathrm{%
of\;mot.})\;,  \label{wz-talg-1} \\
\left\{ \delta ,\delta _\mu \right\}  &=&\partial _\mu \;+(\mathrm{gauge}\;%
\mathrm{transf.)\;}+\;(\mathrm{eqs.}\;\mathrm{of\;mot.})\;,  \nonumber \\
\left\{ \delta _\mu ,\delta _\nu \right\}  &=&(\mathrm{gauge}\;\mathrm{%
transf.)\;}+\;(\mathrm{eqs.}\;\mathrm{of\;mot.})\;,  \nonumber
\end{eqnarray}
and
\end{remark}

\begin{eqnarray}
\left\{ \delta ,\delta _{\mu \nu }\right\} &=&\left\{ \delta _{\mu \nu
},\delta _{\rho \sigma }\right\} =(\mathrm{gauge}\;\mathrm{transf.)}+(%
\mathrm{eqs.}\;\mathrm{of\;mot.})\;,  \label{wz-talg-2} \\
\left\{ \delta _\mu ,\delta _{\rho \sigma }\right\} &=&-(\varepsilon _{\mu
\rho \sigma \nu }\partial ^\nu +g_{\mu \rho }\partial _\sigma -g_{\mu \sigma
}\partial _\rho )+(\mathrm{g.}\;\mathrm{tr.)}+(\mathrm{eqs.\;mot.})\;, 
\nonumber
\end{eqnarray}
As we shall see in the next subsections, it is precisely this twisted
version of the $N=2$\ susy algebra\ which shall be found in the TYM theory.

\subsection{Relationship between TYM\ and N=2 pure Yang-Mills}

Having discussed the twisting procedure, let us now turn to the relationship
between Witten's TYM and the $N=2$ Yang-Mills theory. Let us show, in
particular, that TYM has the same field content of the $N=2$ Yang-Mills
theory in the Wess-Zumino gauge. The minimal $N=2$ supersymmetric\ pure
Yang-Mills theory is described by a gauge multiplet\ which, in the
Wess-Zumino gauge, contains \textit{\cite{magg,mar}}: a gauge field $A_\mu $%
, two spinors $\psi _\alpha ^i\;i=1,2$, and their conjugate $\overline{\psi }%
_{\dot{\alpha}}^i$, two scalars $\phi ,\overline{\phi }\;(\overline{\phi }$
being the complex conjugate of $\phi )$. All these fields are in the adjoint
representations of the gauge group. We also recall that in the Wess-Zumino
gauge the generators $\left( \mathcal{Q}_{\;\alpha }^i,\overline{\mathcal{Q}}%
_{\;\dot{\alpha}}^j\right) $ of $N=2$ act nonlinearly and the supersymmetry
algebra is that of eqs.$\left( \ref{wz-n=2}\right) $.

Let us proceed by applying the previous twisting procedure to the $N=2\;$%
Wess-Zumino gauge multiplet $(A_\mu ,\psi _\alpha ^i,\overline{\psi }_{\dot{%
\alpha}}^i,\phi ,\overline{\phi })$.\ Identifying then the internal index $i$
with the spinor index $\alpha $, it is very easy to see that the spinor $%
\overline{\psi }_{\dot{\alpha}}^i\;$can be related to an anticommuting
vector $\psi _\mu ,$\ i.e

\begin{equation}
\overline{\psi }_{\dot{\alpha}}^i\stackrel{twist}{\longrightarrow }\overline{%
\psi }_{\alpha \dot{\alpha}}\longrightarrow \psi _\mu =(\overline{\sigma }%
_\mu )^{\dot{\alpha}\alpha }\overline{\psi }_{\alpha \dot{\alpha}}\;.
\label{t-spinor}
\end{equation}
Concerning now the fields $\psi _\beta ^i\;$we have

\begin{equation}
\psi _\beta ^i\stackrel{twist}{\longrightarrow }\psi _{\alpha \beta }=\psi
_{(\alpha \beta )}+\psi _{[\alpha \beta ]}\;,  \label{bipsi}
\end{equation}
$\psi _{(\alpha \beta )}\;$and $\psi _{[\alpha \beta ]}\;$being respectively
symmetric and antisymmetric in the spinor indices $\alpha ,\beta $. To the
antisymmetric component $\psi _{[\alpha \beta ]}\;$we associate an
anticommuting scalar field $\eta $

\begin{equation}
\psi _{[\alpha \beta ]}\longrightarrow \eta =\varepsilon ^{\alpha \beta
}\psi _{[\alpha \beta ]}\;,  \label{eta-field}
\end{equation}
while the symmetric part $\psi _{(\alpha \beta )}\;$turns out to be related
to an antisymmetric self-dual field $\chi _{\mu \nu }\;$through\ 

\begin{equation}
\psi _{(\alpha \beta )}\longrightarrow \chi _{\mu \nu }=\widetilde{\chi }%
_{\mu \nu }=(\sigma _{\mu \nu })^{\alpha \beta }\psi _{(\alpha \beta )}\;.
\label{chi-field}
\end{equation}
Therefore, the twisting procedure allows to replace the $N=2\;$Wess-Zumino
multiplet $(A_\mu ,\psi _\alpha ^i,\overline{\psi }_{\dot{\alpha}}^i,\phi ,%
\overline{\phi })$ by the twisted multiplet $(A_\mu ,\psi _\mu ,\chi _{\mu
\nu },\eta ,\phi ,\overline{\phi })$ whose field content is precisely that
of the TYM\ action $\left( \ref{tym}\right) $.

As one can now easily guess, the same property holds for the $N=2$\ pure
Yang-Mills action, as it has been detailed analysed in the Chapters 3 and 6
of \textit{\cite{mar}}, i.e.

\begin{equation}
\mathcal{S}_{YM}^{N=2}(A_\mu ,\psi _\alpha ^i,\overline{\psi }_{\dot{\alpha}%
}^i,\phi ,\overline{\phi })\stackrel{twist}{\longrightarrow }\mathcal{\;S}%
_{TYM}(A_\mu ,\psi _\mu ,\chi _{\mu \nu },\eta ,\phi ,\overline{\phi })\;,
\label{n2ym=tym}
\end{equation}
showing thus that the TYM\ is in fact the twisted version of the ordinary $%
N=2$ Yang-Mills in the Wess-Zumino gauge. This important point, already
underlined by Witten in its original work \textit{\cite{tym}}, deserves a
few clarifying remarks in order to make contact with the results on
topological field theories obtained in the recent years.

\begin{remark}
The first observation is naturally related to the existence of further
symmetries of the TYM\ action $\left( \ref{tym}\right) $. In fact, from eq.$%
\left( \ref{n2ym=tym}\right) $\ one can immediately infer that the TYM
should possess the same symmetry content of the $N=2$\ Yang-Mills. We expect
therefore that, according to the previous analysis, the TYM will be left
invariant by a set of additional transformations whose generators $\left(
\delta ,\delta _\mu ,\delta _{\mu \nu }\right) $\ correspond to the twisted $%
N=2$\ supersymmetric charges $\mathcal{Q}_{\;\alpha }^i,\overline{\mathcal{Q}%
}_{\;\dot{\alpha}}^j$\ and fulfill the Wess-Zumino susy algebra $\left( \ref
{wz-talg-1}\right) ,\left( \ref{wz-talg-2}\right) $. It is easy to check
that Witten's fermionic symmetry $\delta _{\mathcal{W}}$\ of eq.$\left( \ref
{d-transf}\right) $\ corresponds to the twisted scalar generator $\delta $.
Concerning now the vector and the tensor invariances $\delta _\mu ,\delta
_{\mu \nu }$, we shall postpone their detailed analysis to the next
Subsect., limiting here to confirm their existence and their relevance
(especially that of $\delta _\mu $) for the quantum analysis.

\vspace{5mm}
\end{remark}

\begin{remark}
The second remark is related to the standard perturbative Feynman diagram
computations. From eq.$\left( \ref{n2ym=tym}\right) $\ it is very tempting
to argue that the values of quantities like the $\beta $-function should be
the same when computed in the ordinary $N=2$\ Yang-Mills and in the twisted
version. After all, at least at the perturbative level, the twisting
procedure has the effect of a linear change of variables on the fields. The
computation of the\ one loop $\beta $-function for TYM\ has indeed been
performed by R.Brooks et al.\textit{\cite{beta}}. As expected, the result
agrees with that of the pure $N=2\;$Yang-Mills. We recall here that the $N=2$%
\ Yang-Mills $\beta $-function$\;$receives contributions only of one loop
order \textit{\cite{beta1}}. It is also wortwhile to mention that, recently,
the algebraic renormalization analysis of $N=2$\ Yang-Mills theory in the
Wess-Zumino gauge has been carried out by N. Maggiore \textit{\cite{magg}}\
who has shown that the model is anomaly free and that there is only one
possible nontrivial local invariant counterterm, corresponding to a possible
renormalization of the unique gauge coupling constant. As we shall see
later, the same conclusion will be reached in the case of TYM.

\vspace{5mm}
\end{remark}

\begin{remark}
The third point is related to the important and intriguing issue of the
cohomological triviality of the TYM theory. Witten's TYM\ theory is commonly
classified as a topological theory of the cohomological type$\;$\textit{\cite
{tf}}, \ meaning that the TYM action can be expressed as a pure BRST\
variation. Of course, this property seems to be in disagreement with the
interpretation of TYM following from the relation $\left( \ref{n2ym=tym}%
\right) $. Nevertheless, we shall prove that there is a way out, allowing us
to clearly establish to what extent we can consider TYM\ as a cohomological
theory. The point here is that the quantization of the TYM action has to be
done by taking into account the full $N=2\;$twisted$\;$supersymmetric
algebra. This step can be easily handled by following the same procedure
proposed by P. White\ \textit{\cite{white} }in the proof of the ultraviolet
finiteness of $N=4$\ Yang-Mills in the Wess-Zumino gauge. As already
remarked, one introduces constant ghosts associated to the nonlinear
supersymmetric transformations. These constant ghosts allow to define an
extended BRST\ operator which turns out to be nilpotent modulo the (matter)
equations of motion, as the operator of eq.$\left( \ref{ext-brst-op}\right) .
$\ Then, the analysis of the renormalizability can be performed along the
lines of the previous Section. Moreover, the perturbative Feynman expansion
is easily seen to be analytic in these constant ghosts, meaning that the
functional space which is acted upon by the BRST operator is that of the
formal power series in the global parameters. It is precisely the
requirement of analyticity in the constant ghosts which yields nontrivial
cohomology classes, as it has been already established by P.White and N.
Maggiore in the cases of the conventional untwisted $N=2,4$ theories \textit{%
\cite{white,magg}}. This means that, as long as the analyticity requirement
is preserved, the theory is nontrivial and can be viewed as an ordinary
supersymmetric field theory, the fields $(\chi _{\mu \nu },\psi _\mu ,\eta
,\phi ,\overline{\phi })$\ being\ interpreted as matter fields. On the other
hand, if the analyticity requirement is relaxed, the theory becomes
cohomologically trivial and we fall into the well known Labastida-Pernici 
\textit{\cite{lp}}\ and Baulieu-Singer$\;$\textit{\cite{bs}}\ formulations.
It is useful to notice here that in these formulations the fields $(\chi
_{\mu \nu },\psi _\mu ,\eta ,\phi ,\overline{\phi })\;$carry a nonvanishing
ghost number and are no longer considered as matter fields. Rather, they are
interpreted as ghost fields. In particular, $\psi _\mu \;$corresponds to the
ghost of the so called topological shift symmetry \textit{\cite{lp,bs}}. It
is an important fact, however, that TYM\ possesses a nontrivial content also
when considered as a cohomological BRST\ trivial theory. Indeed, as shown by
R. Stora et al.\textit{\cite{stora1,stora2},} the relavant cohomology which
characterizes the TYM in the cohomological version is the so called
equivariant cohomology which, unlike the BRST\ cohomology, is found to be
not empty, allowing to recover the original observables proposed by Witten.
Remarkably in half, the two points of view can be shown to be equivalent, as
proven by the authors \textit{\cite{dmpw,bc}}, who have been able to
establish that the equivariant cohomology coincides in fact with the
cohomology of an extended operator, provided analyticity in a suitable
global parameter is required. In the present case, the role of this global
parameter is palyed by the constant global ghosts of $N=2$ supersymmetry,
the analyticity requirement following from perturbation theory. Summarizing,
the TYM theory can be seen either as a conventional field theory
(analyticity in the global ghosts is here demanded) or as a topological
theory of the cohomologycal type. In this latter case one has to remember
that the Witten's observables should belong to the equivariant cohomology.
We will have the opportunity of showing the explicit equivalence of both
points of view in the analysis of the invariant field polynomial $tr(\phi ^2)
$. Let us conclude this remark by emphasizing that the analyticity
requirement, being naturally related to perturbation theory, is more closed
to the present discussion.
\end{remark}

\subsection{The vector supersymmetry}

Let us now complete the previous analysis by showing that the TYM action $%
\left( \ref{tym}\right) $ possesses indeed further nonlinear symmetries
whose anticommutation relations with the Witten's fermionic symmetry $\delta
_{\mathcal{W}}$\ \ yield precisely the twisted $N=2$ susy algebra of eqs.$%
\left( \ref{wz-talg-1}\right) ,\left( \ref{wz-talg-2}\right) $. Let us first
focus on the vector invariance $\delta _\mu $. To this aim we introduce the
following nonlinear transformations

\begin{eqnarray}
\delta _\mu A_\nu &=&\frac 12\chi _{\mu \nu }+\frac 18g_{\mu \nu }\eta \;,
\label{vsusy} \\
\delta _\mu \psi _\nu &=&F_{\mu \nu }-\frac 12F_{\mu \nu }^{+}-\frac
1{16}g_{\mu \nu }[\phi ,\overline{\phi }]\;,  \nonumber \\
\delta _\mu \eta &=&\frac 12D_\mu \overline{\phi }\;,  \nonumber \\
\delta _\mu \chi _{\sigma \tau } &=&\frac 18(\varepsilon _{\mu \sigma \tau
\nu }D^\nu \overline{\phi }+g_{\mu \sigma }D_\tau \overline{\phi }-g_{\mu
\tau }D_\sigma \overline{\phi })\;,  \nonumber \\
\delta _\mu \phi &=&-\psi _\mu \;,  \nonumber \\
\delta _\mu \overline{\phi } &=&0\;,  \nonumber
\end{eqnarray}
and

\[
\stackunder{Table\;2.}{\stackrel{Dim.and\text{ }\mathcal{R}-charges}{
\begin{tabular}{|c|c|}
\hline
& $\delta _\mu $ \\ \hline
$\mathrm{dim}.$ & $1/2$ \\ \hline
$\mathcal{R}-\mathrm{charg}.$ & $-1$ \\ \hline
$\mathrm{nature}$ & $ant.$ \\ \hline
\end{tabular}
}} 
\]
Transformations $\left( \ref{vsusy}\right) $ are found to leave the TYM
action $\left( \ref{tym}\right) $ invariant

\begin{equation}
\delta _\mu \mathcal{S}_{TYM}=0\;.  \label{vsusy-inv}
\end{equation}
In addition, it is easily verified that the vector generator $\delta _\mu \;$%
gives rise, together with the operator $\delta _{\mathcal{W}}$, to the
following algebraic relations

\begin{eqnarray}
\left\{ \delta _{\mathcal{W}},\delta _\mu \right\} &=&\partial _\mu +\delta
_{A_\mu }^g+(\mathrm{matt.\;eqs.\;of\;motion})\;,  \label{d-vd-alg} \\
\left\{ \delta _\mu ,\delta _\nu \right\} &=&-\frac 18g_{\mu \nu }\delta _{%
\overline{\phi }}^g\;+(\mathrm{matt.\;eqs.\;of\;motion})\;,  \nonumber
\end{eqnarray}
where $\delta _{A_\mu }^g\;$and $\delta _{\overline{\phi }}^g\;$are gauge
transformations with field dependent parameters $A_\mu \;$and $\overline{%
\phi }$, respectively. It is apparent thus from eqs.$\left( \ref{d2}\right)
,\left( \ref{d-vd-alg}\right) $ that, as expected, the operators $\delta _{%
\mathcal{W}}\;$and $\delta _\mu \;$obey the twisted $N=2\;$supersymmetric
algebra $\left( \ref{wz-talg-1}\right) .$

Let us also write down, for further use, the explicit form of eqs.$\left( 
\ref{d-vd-alg}\right) $,\ \textit{i.e.}

\begin{eqnarray}
\left\{ \delta _\mu ,\delta _\nu \right\} A_\sigma &=&\frac 18g_{\mu \nu
}D_\sigma \overline{\phi }\;,  \label{d-vd-alg-A} \\
\left\{ \delta _{\mathcal{W}},\delta _\mu \right\} A_\sigma &=&\partial _\mu
A_\sigma -D_\sigma A_\mu \;,  \nonumber
\end{eqnarray}

\begin{eqnarray}
\left\{ \delta _{\mathcal{W}},\delta _\mu \right\} \psi _\sigma &=&\partial
_\mu \psi _\sigma +[A_\mu ,\psi _\sigma ]\;+\frac{g^2}4\frac{\delta \mathcal{%
S}_{TYM}}{\delta \chi ^{\mu \sigma }}\;,\;  \label{d-vd-alg-psi} \\
\left\{ \delta _\mu ,\delta _\nu \right\} \psi _\sigma &=&-\frac 18g_{\mu
\nu }[\overline{\phi },\psi _\sigma ]+\frac{g^2}{16}\left( g_{\mu \sigma
}g_{\nu \tau }+g_{\nu \sigma }g_{\mu \tau }-2g_{\mu \nu }g_{\sigma \tau
}\right) \frac{\delta \mathcal{S}_{TYM}}{\delta \psi _\tau }\;,  \nonumber
\end{eqnarray}

\begin{eqnarray}
\left\{ \delta _\mu ,\delta _\nu \right\} \eta &=&-\frac 18g_{\mu \nu }[%
\overline{\phi },\eta ]\;,  \label{d-vd-alg-eta} \\
\left\{ \delta _{\mathcal{W}},\delta _\mu \right\} \eta &=&\partial _\mu
\eta +[A_\mu ,\eta ]\;,  \nonumber
\end{eqnarray}

\begin{eqnarray}
\left\{ \delta _\mu ,\delta _\nu \right\} \chi _{\sigma \tau } &=&-\frac
18g_{\mu \nu }[\overline{\phi },\chi _{\sigma \tau }]\;,
\label{d-vd-alg-chi} \\
\left\{ \delta _{\mathcal{W}},\delta _\mu \right\} \chi _{\sigma \tau }
&=&\partial _\mu \chi _{\sigma \tau }+[A_\mu ,\chi _{\sigma \tau }]+\frac{g^2%
}8\left( \varepsilon _{\mu \sigma \tau \nu }+g_{\mu \sigma }g_{\tau \nu
}-g_{\mu \tau }g_{\sigma \nu }\right) \frac{\delta \mathcal{S}_{TYM}}{\delta
\psi _\nu }\;,  \nonumber
\end{eqnarray}

\begin{eqnarray}
\;\left\{ \delta _\mu ,\delta _\nu \right\} \phi &=&-\frac 18g_{\mu \nu }[%
\overline{\phi },\phi ]\;,  \label{d-vd-alg-phi} \\
\left\{ \delta _{\mathcal{W}},\delta _\mu \right\} \phi &=&\partial _\mu
\phi +[A_\mu ,\phi ]\;,  \nonumber
\end{eqnarray}
and 
\begin{eqnarray}
\left\{ \delta _\mu ,\delta _\nu \right\} \overline{\phi } &=&0\;,
\label{d-vd-alg-phib} \\
\left\{ \delta _{\mathcal{W}},\delta _\mu \right\} \overline{\phi }
&=&\partial _\mu \overline{\phi }+[A_\mu ,\overline{\phi }]\;.  \nonumber
\end{eqnarray}

\begin{remark}
We underline here that the form of the TYM action $\left( \ref{tym}\right) $%
, not completely specified by the fermionic symmetry$\;\delta _{\mathcal{W}}$%
, turns out to be uniquely characterized by the vector invariance $\delta
_\mu $. In other words, eqs.$\left( \ref{vsusy-inv}\right) $ and $\left( \ref
{d-vd-alg}\right) $ fix all the relative numerical coefficients of the
Witten's action $\left( \ref{tym}\right) $ allowing, in particular, for a
single coupling constant. This feature will be of \ great importance for the
renormalizability analysis of the model.

\vspace{5mm}
\end{remark}

Concerning now the existence of a self-dual symmetry $\delta _{\mu \nu },\;$%
we shall remind the reader to the App.B, where the explicit form of the
self-dual transformations$\;$will be given. Needless to say, the self-dual
generator $\delta _{\mu \nu }$ will reproduce, together with the operators $%
\delta _{\mathcal{W}},\delta _\mu $, the complete $N=2\;$susy algebra $%
\left( \ref{wz-talg-1}\right) ,\left( \ref{wz-talg-2}\right) $. The reasons
why we do not actually take in further account the self-dual transformations 
$\delta _{\mu \nu }$ are due partly to the fact that, as previously
remarked, the TYM action is already uniquely fixed by the$\;(\delta _{%
\mathcal{W}}$, $\delta _\mu )-$symmetries and partly to the fact that the
generator $\delta _{\mu \nu }\;$turns out to be almost trivially realized on
the fields, as one can easily infer from the App.B. Looking for instance at
the $\delta _{\mu \nu }$-transformations of the fields $A_\sigma \;$and\ $%
\psi _\sigma $, it is apparent to check that they can be rewritten as

\begin{eqnarray}
\delta _{\mu \nu }A_\sigma &=&-(\varepsilon _{\mu \nu \sigma \tau }+g_{\mu
\sigma }g_{\nu \tau }-g_{\nu \sigma }g_{\mu \tau })\delta _{\mathcal{W}%
}A^\tau \;,  \label{triv-real} \\
\delta _{\mu \nu }\psi _\sigma &=&\;(\varepsilon _{\mu \nu \sigma \tau
}+g_{\mu \sigma }g_{\nu \tau }-g_{\nu \sigma }g_{\mu \tau })\delta _{%
\mathcal{W}}\psi ^\tau \;,  \nonumber
\end{eqnarray}
showing in fact that the $\delta _{\mu \nu }$-transformations can be
trivially realized in terms of $\delta _{\mathcal{W}}$-transformations.$\;$%
This means that the subalgebra $\left( \ref{wz-talg-1}\right) $ carries
essentially al the relevant informations concerning the $N=2\;$%
supersymmetric structure ot the TYM.

We can turn now to the quantization of the model. This will be the task of
the next Section.

\begin{remark}
Of course, all the nonlinear $(\delta _{\mathcal{W}},\delta _\mu ,\delta
_{\mu \nu })$-transformations$\;$of the fields of TYM can be obtained by
performing the twist of the conventional ($\mathcal{Q}_{\;\alpha }^i,%
\overline{\mathcal{Q}}_{\;\dot{\alpha}}^j$)-transformations of the untwisted
N=2 susy Wess-Zumino multiplet $(A_\mu ,\psi _\alpha ^i,\overline{\psi }_{%
\dot{\alpha}}^i,\phi ,\overline{\phi })$.
\end{remark}

\section{\ Quantizing topological Yang-Mills}

\subsection{Extendend BRST operator and gauge fixing}

As we have seen in the previous Section, the TYM\ action $\mathcal{S}%
_{TYM}\; $is left invariant by a set of nonlinear symmetries whose
generators $\delta _{\mathcal{W}}$, $\delta _\mu $, give rise to an algebra $%
\left( \ref{wz-talg-1}\right) $ of the type of that considered in eq.$\left( 
\ref{add-algebra}\right) $. Following therefore the discussion of Sect.2.,
we shall begin by looking at an extended BRST operator $\mathcal{Q\;}$which
turns out to be nilpotent on shell. To this purpose we first introduce the
Faddeev-Popov ghost field $c\;$corresponding to the local gauge invariance $%
\left( \ref{g-inv}\right) $ of the action $\left( \ref{tym}\right) $,

\begin{equation}
\epsilon \rightarrow c\;,\;\delta _\epsilon ^g\rightarrow s\;,  \label{s-op}
\end{equation}
with

\begin{eqnarray}
sA_\mu &=&-D_\mu c\;,  \label{s-op=transf} \\
s\psi _\mu &=&\left\{ c,\psi _\mu \right\} \;,  \nonumber \\
s\chi _{\mu \nu } &=&\left\{ c,\chi _{\mu \nu }\right\} \;,  \nonumber \\
s\eta &=&\left\{ c,\eta \right\} \;,  \nonumber \\
s\phi &=&\left[ c,\phi \right] \;,  \nonumber \\
s\overline{\phi } &=&\left[ c,\overline{\phi }\right] \;,  \nonumber \\
sc &=&c^2\;, \\
s^2 &=&0\;,  \nonumber
\end{eqnarray}
and

\begin{equation}
s\mathcal{S}_{TYM}=0\;.  \label{s-op-inv}
\end{equation}
We associate now to each generator entering the algebra $\left( \ref
{wz-talg-1}\right) $, namely $\delta _{\mathcal{W}}$, $\delta _\mu \;$and $%
\partial _\mu $, the corresponding\ constant ghost parameters $\left( \omega
,\varepsilon ^\mu ,v^\mu \right) $

\begin{equation}
\omega \rightarrow \delta _{\mathcal{W}}\;,\;\;\;\varepsilon ^\mu
\rightarrow \delta _\mu \;,\;\;\;v^\mu \rightarrow \partial _\mu \;.\;
\label{c-g-p}
\end{equation}
with

\[
\stackunder{Table\;3.}{\stackrel{Dim.,\text{ }\mathcal{R}-charges\;and%
\;gh-numbers}{
\begin{tabular}{|c|c|c|c|c|}
\hline
& $c$ & $\omega $ & $\varepsilon ^\mu $ & $v^\mu $ \\ \hline
$\dim .$ & $0$ & $-1/2$ & $-1/2$ & $-1$ \\ \hline
$\mathcal{R-}\mathrm{ch}\arg \mathrm{e}$ & $0$ & $-1$ & $1$ & $0$ \\ \hline
$\mathrm{gh-number}$ & $1$ & $1$ & $1$ & $1$ \\ \hline
$\mathrm{nature}$ & $ant.$ & $comm.$ & $comm.$ & $ant.$ \\ \hline
\end{tabular}
}} 
\]
Therefore, the extended BRST operator

\begin{equation}
\mathcal{Q=}s+\omega \delta _{\mathcal{W}}+\varepsilon ^\mu \delta _\mu
+v^\mu \partial _\mu -\omega \varepsilon ^\mu \frac \partial {\partial v^\mu
}\;,  \label{Q-op}
\end{equation}
has ghost number one, vanishing $\mathcal{R}$-charge, and

\begin{eqnarray}
\mathcal{QS}_{TYM} &=&0\;,  \label{QQ} \\
\mathcal{Q}^2 &=&(\text{\textrm{matter eqs. of mot.}})\;.  \nonumber
\end{eqnarray}

\begin{remark}
We underline here that the fields $(A_\mu ,\psi _\mu ,\eta ,\chi _{\mu \nu
},\phi ,\overline{\phi })$ do not carry any ghost number. In particular $%
(\psi _\mu ,\eta ,\chi _{\mu \nu },\phi ,\overline{\phi })\;$are considered
as (twisted) matter fields.
\end{remark}

\vspace{5mm}

While the first condition of the above equation follows from the simple
observation that the TYM action $\left( \ref{tym}\right) $ does not depend
from the ghosts and that it is invariant under ordinary space-time
translations, the second one requires some care and follows from defining in
a suitable way the action of the four generators $s,\delta _{\mathcal{W}%
},\delta _\mu $ and $\partial _\mu $\ on the ghosts $\left( c,\omega
,\varepsilon ^\mu ,v^\mu \right) $. To have a more precise idea of how this
is done, let us work out in detail the case of the two operators $s$\ and $%
\delta _{\mathcal{W}}$. Recalling that $\;$ 
\begin{equation}
\delta _{\mathcal{W}}^2=\delta _\phi ^g\;+\;(\chi \text{\textrm{-eq. of
motion}})\;,  \label{d2-b}
\end{equation}
one looks then for an operator $(s+\omega \delta _{\mathcal{W}})$ such that

\begin{eqnarray}
(s+\omega \delta _{\mathcal{W}})^2 &=&\;0\;\;\;\mathrm{on\;\;\;}(A_\mu ,\psi
_\mu ,\eta ,\phi ,\overline{\phi },c,\omega )\;,  \label{s-om-d} \\
(s+\omega \delta _{\mathcal{W}})^2\chi _{\mu \nu } &=&(\chi \text{\textrm{%
-eq. of motion}})\;.  \nonumber
\end{eqnarray}
After a little experiment, it is not difficult to convince oneself that the
above conditions are indeed verified by defining the action of $s$\ and $%
\delta _{\mathcal{W}}\;$on the ghost $(c,\omega )\;$as

\begin{equation}
s\omega =0\;,\;\;\;\delta _{\mathcal{W}}\omega =0\;,\;\;\;\delta _{\mathcal{W%
}}c=-\omega \phi \;.  \label{s-om-d-def}
\end{equation}
Notice that the only nontrivial extension is that of the operator $\delta _{%
\mathcal{W}}\;$on the Faddeev-Popov ghost $c$. It is precisely this
transformation which compensates the gauge transformation $\delta _\phi ^g\;$%
in the right hand side of eq.$\left( \ref{d2-b}\right) $, ensuring then the
on shell nilpotency of the operator $(s+\omega \delta _{\mathcal{W}})$.

The above procedure can be now easily repeated in order to include in the
game also the operators $\delta _\mu $ and $\partial _\mu $. The final
result is that the extension of the operator $\mathcal{Q}$ on the ghosts $%
\left( c,\omega ,\varepsilon ^\mu ,v^\mu \right) $ is found to be

\begin{eqnarray}
\mathcal{Q}c &=&c^2-\omega ^2\phi -\omega \varepsilon ^\mu A_\mu +\frac{%
\varepsilon ^2}{16}\overline{\phi }+v^\mu \partial _\mu c\;,  \label{Q-ext}
\\
\mathcal{Q}\omega &=&0\;,\;\;\;\mathcal{Q}\varepsilon ^\mu =0\;,  \nonumber
\\
\mathcal{Q}v^\mu &=&-\omega \varepsilon ^\mu \;.  \nonumber
\end{eqnarray}
The construction of the gauge fixing term is now almost trivial. We
introduce an antighost $\overline{c}$ and a Lagrangian multiplier $b$
transforming as \textit{\cite{white,magg}}$\;$%
\begin{eqnarray}
\mathcal{Q}\overline{c}\; &=&b+v^\mu \partial _\mu \overline{c}\;,
\label{cb-b} \\
\mathcal{Q}b &=&\omega \varepsilon ^\mu \partial _\mu \overline{c}+v^\mu
\partial _\mu b\;,  \nonumber
\end{eqnarray}
with

\begin{equation}
\mathcal{Q}^2\overline{c}=\mathcal{Q}^2b=0\;.  \label{Q-cb-b}
\end{equation}

\[
\stackunder{Table\;4.}{\stackrel{Dim.,\text{ }\mathcal{R}-charges\;and%
\;gh-numbers}{
\begin{tabular}{|c|c|c|}
\hline
& $\overline{c}$ & $b$ \\ \hline
$\dim $ & $2$ & $2$ \\ \hline
$\mathcal{R-}\mathrm{ch\arg e}$ & $0$ & $0$ \\ \hline
$\mathrm{gh-number}$ & $-1$ & $0$ \\ \hline
$\mathrm{nature}$ & $ant.$ & $comm.$ \\ \hline
\end{tabular}
}} 
\]
According to the eq.$\left( \ref{gauge-fix}\right) $, for the gauge fixing
action we thus get\ 

\begin{eqnarray}
S_{gf} &=&Q\int d^4x\;tr(\overline{c}\partial A)  \label{landau-g-fi} \\
&=&tr\int d^4x\;\left( b\partial ^\mu A_\mu +\overline{c}\partial ^\mu D_\mu
c-\omega \overline{c}\partial ^\mu \psi _\mu -\frac{\varepsilon ^\nu }2%
\overline{c}\partial ^\mu \chi _{\nu \mu }-\frac{\varepsilon ^\mu }8%
\overline{c}\partial _\mu \eta \right) \;,  \nonumber
\end{eqnarray}
so that the gauge fixed action $\left( \mathcal{S}_{TYM}+S_{gf}\right) \;$is 
$\mathcal{Q}$-invariant,

\begin{equation}
\mathcal{Q}\left( \mathcal{S}_{TYM}+S_{gf}\right) =0\;.  \label{Q-g-f-action}
\end{equation}
The above equation means that the gauge fixing procedure has been worked out
by taking into account not only the pure local gauge symmetry but also the
additional nonlinear invariances $\delta _{\mathcal{W}}\;$and $\delta _\mu $%
, as one can easily deduce from the explicit dependence of the gauge fixing
term $\left( \ref{landau-g-fi}\right) $ from the global ghosts $\omega
,\varepsilon ^\mu $. Of course, the absence of the ghost $v^\mu \;$is due to
the space-time translation invariance of expression $\left( \ref{landau-g-fi}%
\right) $.

Let us conclude this paragraph by summarizing all the properties of the
extended operator $\mathcal{Q}$, \textit{i.e}.

\begin{eqnarray}
\mathcal{Q}A_\mu &=&-D_\mu c+\omega \psi _\mu +\frac{\varepsilon ^\nu }2\chi
_{\nu \mu }+\frac{\varepsilon _\mu }8\eta +v^\nu \partial _\nu A_\mu \;,
\label{Q-transf} \\
\mathcal{Q}\psi _\mu &=&\left\{ c,\psi _\mu \right\} -\omega D_\mu \phi
+\varepsilon ^\nu \left( F_{\nu \mu }-\frac 12F_{\nu \mu }^{+}\right) -\frac{%
\varepsilon _\mu }{16}[\phi ,\overline{\phi }]  \nonumber \\
&&+v^\nu \partial _\nu \psi _\mu \;,  \nonumber \\
\mathcal{Q}\chi _{\sigma \tau } &=&\left\{ c,\chi _{\sigma \tau }\right\}
+\omega F_{\sigma \tau }^{+}+\frac{\varepsilon ^\mu }8(\varepsilon _{\mu
\sigma \tau \nu }+g_{\mu \sigma }g_{\nu \tau }-g_{\mu \tau }g_{\nu \sigma
})D^\nu \overline{\phi }\;  \nonumber \\
&&+v^\nu \partial _\nu \chi _{\sigma \tau }\;,  \nonumber \\
\mathcal{Q}\eta &=&\left\{ c,\eta \right\} +\frac \omega 2[\phi ,\overline{%
\phi }]+\frac{\varepsilon ^\mu }2D_\mu \overline{\phi }+v^\nu \partial _\nu
\eta \;,  \nonumber \\
\mathcal{Q}\phi &=&\left[ c,\phi \right] -\varepsilon ^\mu \psi _\mu +v^\nu
\partial _\nu \phi \;,  \nonumber \\
\mathcal{Q}\overline{\phi } &=&\left[ c,\overline{\phi }\right] +2\omega
\eta +v^\nu \partial _\nu \overline{\phi }\;,  \nonumber \\
\mathcal{Q}c &=&c^2-\omega ^2\phi -\omega \varepsilon ^\mu A_\mu +\frac{%
\varepsilon ^2}{16}\overline{\phi }+v^\nu \partial _\nu c\;,  \nonumber \\
\mathcal{Q}\omega &=&0\;,\;\;\;  \nonumber \\
\mathcal{Q}\varepsilon ^\mu &=&0\;,\;\;\;  \nonumber \\
\mathcal{Q}v^\mu &=&-\omega \varepsilon ^\mu \;,  \nonumber \\
\mathcal{Q}\overline{c}\; &=&b+v^\mu \partial _\mu \overline{c}\;,\; 
\nonumber \\
\mathcal{Q}b &=&\omega \varepsilon ^\mu \partial _\mu \overline{c}+v^\mu
\partial _\mu b\;,  \nonumber
\end{eqnarray}
with

\begin{equation}
\mathcal{Q}^2=0\;\;\;\;\mathrm{on\;\;\;\;}\left( A,\phi ,\overline{\phi }%
,\eta ,c,\omega ,\varepsilon ,v,\overline{c},b\right) \;,  \label{QQ-1}
\end{equation}
and

\begin{eqnarray}
\mathcal{Q}^2\psi _\sigma &=&\frac{g^2}4\omega \varepsilon ^\mu \frac{\delta 
\mathcal{S}_{TYM}}{\delta \chi ^{\mu \sigma }}  \label{QQ-2} \\
&&+\frac{g^2}{32}\varepsilon ^\mu \varepsilon ^\nu \left( g_{\mu \sigma }%
\frac{\delta \mathcal{S}_{TYM}}{\delta \psi ^\nu }+g_{\nu \sigma }\frac{%
\delta \mathcal{S}_{TYM}}{\delta \psi ^\mu }-2g_{\mu \nu }\frac{\delta 
\mathcal{S}_{TYM}}{\delta \psi ^\sigma }\right) \;,  \nonumber
\end{eqnarray}

\begin{eqnarray}
\mathcal{Q}^2\chi _{\sigma \tau } &=&-\frac{g^2}2\omega ^2\frac{\delta 
\mathcal{S}_{TYM}}{\delta \chi ^{\sigma \tau }}  \label{QQ-3} \\
&&+\frac{g^2}8\omega \varepsilon ^\mu \left( \varepsilon _{\mu \sigma \tau
\nu }\frac{\delta \mathcal{S}_{TYM}}{\delta \psi _\nu }+g_{\mu \sigma }\frac{%
\delta \mathcal{S}_{TYM}}{\delta \psi ^\tau }-g_{\mu \tau }\frac{\delta 
\mathcal{S}_{TYM}}{\delta \psi ^\sigma }\right) \;.  \nonumber
\end{eqnarray}

\begin{remark}
Notice that the $\mathcal{Q}$-transformation of the Faddeev-Popov ghost $c$
contains terms quadratic in the global parameters $\omega ,\varepsilon ^\mu $%
. The presence of these terms (in particular of $\omega ^2\phi $) in the
transformation of the ghost $c$ has been shown to be of great importance by
the authors of ref. \textit{\cite{dmpw}} in order to identify the relevant
nontrivial cohomology classes of TYM.\ \ 
\end{remark}

\subsection{The Slavnov-Taylor identity}

As explained in Sect.2, in order to obtain the Slavnov-Taylor identity\ we
first couple the nonlinear $\mathcal{Q}$-transformations of the fields in
eqs.$\left( \ref{Q-transf}\right) $ to a set of antifields $(L,D,\Omega ^\mu
,\xi ^\mu ,\rho ,\tau ,B_{\mu \nu }),$

\begin{eqnarray}
\mathcal{S}_{ext} &=&\displaystyle tr\int d^4x\;(\;L\mathcal{Q}c+D\mathcal{Q}%
\phi +\Omega ^\mu \mathcal{Q}A_\mu +\xi ^\mu \mathcal{Q}\psi _\mu \;
\label{s-ext} \\
&&\;\;\;\;\;\;\;\;\;\;\;\;\;\;\;\;\;\;\;+\rho \mathcal{Q}\overline{\phi }%
+\tau \mathcal{Q}\eta +\frac 12B^{\mu \nu }\mathcal{Q\chi }_{\mu \nu }\;)\;,
\nonumber
\end{eqnarray}
with

\[
\stackunder{Table\;5.}{\stackrel{Dim.,\;\mathcal{R}-charges\;and\;gh-numbers%
}{
\begin{tabular}{|c|c|c|c|c|c|c|c|}
\hline
& $L$ & $D$ & $\Omega ^\mu $ & $\xi ^\mu $ & $\rho $ & $\tau $ & $B^{\mu \nu
}$ \\ \hline
$\dim .$ & $4$ & $3$ & $3$ & $5/2$ & $3$ & $5/2$ & $5/2$ \\ \hline
$\mathcal{R}-\mathrm{ch}\arg \mathrm{e}$ & $0$ & $-2$ & $0$ & $-1$ & $2$ & $%
1 $ & $1$ \\ \hline
$\mathrm{gh-number}$ & $-2$ & $-1$ & $-1$ & $-1$ & $-1$ & $-1$ & $-1$ \\ 
\hline
$\mathrm{nature}$ & $comm.$ & $ant.$ & $ant.$ & $comm.$ & $ant.$ & $comm.$ & 
$comm.$ \\ \hline
\end{tabular}
}} 
\]
Moreover, taking into account that the extended operator $\mathcal{Q}$ is
nilpotent only modulo the equations of motion of the fields $\psi _\mu $ and 
$\;\chi _{\mu \nu }$, we also introduce a term quadratic in the
corresponding antifields $\xi ^\mu ,\;B^{\mu \nu },\;$\textit{i.e.}

\begin{equation}
\mathcal{S}_{quad}=tr\int d^4x\left( \frac \alpha 4\omega ^2B^{\mu \nu
}B_{\mu \nu }+\frac \beta 2\omega B^{\mu \nu }\varepsilon _\mu \xi _\nu
+\frac \lambda 2\varepsilon ^\mu \varepsilon ^\nu \xi _\mu \xi _\nu +\frac
\gamma 2\varepsilon ^2\xi ^2\right) \;,  \label{s-quad}
\end{equation}
where the coefficients $(\alpha ,\beta ,\lambda ,\gamma )\;$are fixed by
requiring that the complete action

\begin{equation}
\Sigma =\mathcal{S}_{TYM}+S_{gf}+\mathcal{S}_{ext}+\mathcal{S}_{quad}\;,
\label{c-action}
\end{equation}
obeys the following identity

\begin{equation}
\mathcal{S}(\Sigma )\;\mathcal{=\;}0\;,  \label{tym-s-t}
\end{equation}
with

\begin{eqnarray}
\mathcal{S}(\Sigma ) &=&tr\int d^4x\left( \frac{\delta \Sigma }{\delta A^\mu 
}\frac{\delta \Sigma }{\delta \Omega _\mu }+\frac{\delta \Sigma }{\delta \xi
^\mu }\frac{\delta \Sigma }{\delta \psi _\mu }+\frac{\delta \Sigma }{\delta L%
}\frac{\delta \Sigma }{\delta c}+\frac{\delta \Sigma }{\delta D}\frac{\delta
\Sigma }{\delta \phi }\;+\frac{\delta \Sigma }{\delta \rho }\frac{\delta
\Sigma }{\delta \overline{\phi }}\right.  \nonumber  \label{tym-s-t-op} \\
&&\;\;\;\;\;\;\;\;\;\;\;\;\;\;+\frac{\delta \Sigma }{\delta \tau }\frac{%
\delta \Sigma }{\delta \eta }+\frac 12\frac{\delta \Sigma }{\delta B^{\mu
\nu }}\frac{\delta \Sigma }{\delta \chi _{\mu \nu }}\;\;+(b+v^\mu \partial
_\mu \overline{c})\frac{\delta \Sigma }{\delta \overline{c}}  \nonumber \\
&&\;\;\;\;\;\;\;\;\;\;\;\;\;\;\left. +(\omega \varepsilon ^\mu \partial _\mu 
\overline{c}+v^\mu \partial _\mu b)\frac{\delta \Sigma }{\delta b}\right)
\;-\omega \varepsilon ^\mu \frac{\partial \Sigma }{\partial v^\mu }\;.
\label{tym-s-t-op}
\end{eqnarray}
The condition $\left( \ref{tym-s-t}\right) $ is easily worked out, yielding
for the coefficients $\alpha ,$ $\beta ,$ $\lambda $ and $\gamma $ the
following values

\begin{equation}
\alpha =\frac{g^2}2,\;\;\;\;\beta =-\frac{g^2}2,\;\;\;\;\lambda =-\frac{g^2}{%
16},\;\;\;\;\gamma =\frac{g^2}{16}\;.  \label{coeffic.}
\end{equation}
The equation $\left( \ref{tym-s-t}\right) \;$yields thus the classical
Slavnov-Taylor identity for the TYM\ and will be the starting point for the
analysis of the renormalizability of the model. However, before entering
into the quantum aspects, let us make some further useful considerations
which allow to cast the Slavnov-Taylor identity $\left( \ref{tym-s-t}\right)
\;$in a simplified form which is more suitable for the quantum discussion.

\subsection{Analysis of the classical Slavnov-Taylor identity}

In order to obtain a simplified version of the Slavnov-Taylor identity, we
shall make use of the fact that the complete action $\Sigma \;$is invariant
under space-time translations, as expressed by

\begin{eqnarray}
\mathcal{P}_\mu \Sigma &=&\stackunder{i}{\sum }\int d^4x\left( \partial _\mu
\varphi ^i\frac{\delta \Sigma }{\delta \varphi ^i}+\partial _\mu \varphi
^{*i}\frac{\delta \Sigma }{\delta \varphi ^{*i}}\right) =\;0\;,\;\;
\label{transl-inv} \\
\;\;\varphi ^i &=&\mathrm{all\;the\;fields\;(}A,\psi ,\phi ,\overline{\phi }%
,\eta ,\chi ,c,\overline{c},b\mathrm{)\;},  \nonumber \\
\varphi ^{*i} &=&\mathrm{all\;the\;antifields\;(}\Omega ,\xi ,L,D,\rho ,\tau
,B\mathrm{)}\;.
\end{eqnarray}
Let us now observe that, as a consequence of the fact that $\mathcal{P}_\mu
\;$acts linearly on the fields and antifields, the dependence of the
complete action $\Sigma \;$from the corresponding translation constant ghost 
$v^\mu \;$turns out to be fixed by the following linearly broken Ward
identity, namely

\begin{equation}
\frac{\partial \Sigma }{\partial v^\mu }=\Delta _\mu ^{cl}\;,  \label{v-id}
\end{equation}
where

\begin{eqnarray}
\Delta _\mu ^{cl} &=&tr\int d^4x(\;L\partial _\mu c-D\partial _\mu \phi
-\Omega ^\nu \partial _\mu A_\nu +\xi ^\nu \partial _\mu \psi _\nu \; 
\nonumber \\
&&\;\;\;\;\;\;\;\;\;\;\;\;\;\;\;-\rho \partial _\mu \overline{\phi }+\tau
\partial _\mu \eta +\frac 12B^{\nu \sigma }\partial _\mu \chi _{\nu \sigma
}\;)\;,  \label{v-break}
\end{eqnarray}
is a classical breaking, being linear in the quantum fields. We fall thus in
the situation described in the \textbf{Remark 5 }of Sect.2, meaning that we
can completely eliminate the global constant ghost $v^\mu $ without any
further consequence. Introducing in fact the action $\widehat{\Sigma }$
through

\begin{eqnarray}
\Sigma &=&\widehat{\Sigma }+v^\mu \Delta _\mu ^{cl}\;,  \label{h-action} \\
\frac{\partial \widehat{\Sigma }}{\partial v^\mu } &=&0\;,  \nonumber
\end{eqnarray}
it is easily verified from $\left( \ref{tym-s-t}\right) $ that $\widehat{%
\Sigma }$ obeys the modified Slavnov-Taylor identity

\begin{equation}
\mathcal{S}(\widehat{\Sigma })\;\mathcal{=\;}\omega \varepsilon ^\mu \Delta
_\mu ^{cl}\;\;,  \label{n-tym-s-t}
\end{equation}
with

\begin{eqnarray}
\mathcal{S}(\widehat{\Sigma }) &=&tr\int d^4x\left( \frac{\delta \widehat{%
\Sigma }}{\delta A^\mu }\frac{\delta \widehat{\Sigma }}{\delta \Omega _\mu }+%
\frac{\delta \widehat{\Sigma }}{\delta \xi ^\mu }\frac{\delta \widehat{%
\Sigma }}{\delta \psi _\mu }+\frac{\delta \widehat{\Sigma }}{\delta L}\frac{%
\delta \widehat{\Sigma }}{\delta c}+\frac{\delta \widehat{\Sigma }}{\delta D}%
\frac{\delta \widehat{\Sigma }}{\delta \phi }\;+\frac{\delta \widehat{\Sigma 
}}{\delta \rho }\frac{\delta \widehat{\Sigma }}{\delta \overline{\phi }}%
\right.  \nonumber  \label{n-tym-st-op} \\
&&\;\;\;\;\;\;\;\;\;\;\;\;\;\;\;\;\left. +\frac{\delta \widehat{\Sigma }}{%
\delta \tau }\frac{\delta \widehat{\Sigma }}{\delta \eta }+\frac 12\frac{%
\delta \widehat{\Sigma }}{\delta B^{\mu \nu }}\frac{\delta \widehat{\Sigma }%
}{\delta \chi _{\mu \nu }}\;+b\frac{\delta \widehat{\Sigma }}{\delta 
\overline{c}}+\omega \varepsilon ^\mu \partial _\mu \overline{c}\frac{\delta 
\widehat{\Sigma }}{\delta b}\right) \;,  \nonumber  \label{n-tym-st-op} \\
&&  \label{n-tym-st-op}
\end{eqnarray}
and $\Delta _\mu ^{cl}\;$as in eq.$\left( \ref{v-break}\right) $.\ The
equation $\left( \ref{n-tym-s-t}\right) \ $represents the final form of the
Slavnov-Taylor identity which will be taken as the starting point for the
quantum analysis of the model. It is interesting to observe that, due to the
elimination of the ghost parameter $v^\mu $, the classical breaking $\Delta
_\mu ^{cl}\;$appears now on the right hand side of the identity $\left( \ref
{n-tym-s-t}\right) $, yielding thus a linearly broken Slavnov-Taylor
identity. As a consequence the linearized Slavnov-Taylor operator $\mathcal{B%
}_{\widehat{\Sigma }}\;$defined as

\begin{eqnarray}
\mathcal{B}_{\widehat{\Sigma }} &=&tr\int d^4x\left( \frac{\delta \widehat{%
\Sigma }}{\delta A^\mu }\frac \delta {\delta \Omega _\mu }+\frac{\delta 
\widehat{\Sigma }}{\delta \Omega _\mu }\frac \delta {\delta A^\mu }+\frac{%
\delta \widehat{\Sigma }}{\delta \psi _\mu }\frac \delta {\delta \xi ^\mu }+%
\frac{\delta \widehat{\Sigma }}{\delta \xi ^\mu }\frac \delta {\delta \psi
_\mu }+\frac{\delta \widehat{\Sigma }}{\delta L}\frac \delta {\delta
c}\right)  \nonumber  \label{n-tym-lin-op} \\
&&\;\;\;\;\;\;\;\;\;+\frac{\delta \widehat{\Sigma }}{\delta c}\frac \delta
{\delta L}+\frac{\delta \widehat{\Sigma }}{\delta \phi }\frac \delta {\delta
D}+\frac{\delta \widehat{\Sigma }}{\delta D}\frac \delta {\delta \phi }+%
\frac{\delta \widehat{\Sigma }}{\delta \overline{\phi }}\frac \delta {\delta
\rho }+\frac{\delta \widehat{\Sigma }}{\delta \rho }\frac \delta {\delta 
\overline{\phi }}+\frac{\delta \widehat{\Sigma }}{\delta \eta }\frac \delta
{\delta \tau }  \nonumber \\
&&\;\;\;\;\;\;\;\;\;\left. +\frac{\delta \widehat{\Sigma }}{\delta \tau }%
\frac \delta {\delta \eta }+\frac 12\frac{\delta \widehat{\Sigma }}{\delta
\chi _{\mu \nu }}\frac \delta {\delta B^{\mu \nu }}+\frac 12\frac{\delta 
\widehat{\Sigma }}{\delta B^{\mu \nu }}\frac \delta {\delta \chi _{\mu \nu
}}+b\frac \delta {\delta \overline{c}}+\omega \varepsilon ^\mu \partial _\mu 
\overline{c}\frac \delta {\delta b}\right) \;,  \nonumber \\
&&\;\;\;\;\;\;\;\;\;\;\;\;  \label{n-tym-lin-op}
\end{eqnarray}
is not strictly nilpotent. Instead, we have

\begin{equation}
\mathcal{B}_{\widehat{\Sigma }}\mathcal{B}_{\widehat{\Sigma }}=\omega
\varepsilon ^\mu \mathcal{P}_\mu \;,  \label{nil-lin-tym}
\end{equation}
meaning that $\mathcal{B}_{\widehat{\Sigma }}\;$is nilpotent only modulo a
total derivative. It follows then that $\mathcal{B}_{\widehat{\Sigma }}\;$%
becomes a nilpotent operator when acting on the space of the integrated
local polynomials in the fields and antifields. This is the case, for
instance, of the invariant counterterms and of the anomalies.

Besides the Slavnov-Taylor identity $\left( \ref{n-tym-s-t}\right) $, the
classical action $\widehat{\Sigma }\;$turns out to be characterized by
further additional constraints \textit{\cite{book}}, namely

\vspace{5mm}

$\bullet \;$the Landau gauge fixing condition

\begin{equation}
\frac{\delta \widehat{\Sigma }}{\delta b}=\partial A\;,  \label{landau-g-f}
\end{equation}

$\bullet \;$the antighost equation 
\begin{equation}
\frac{\delta \widehat{\Sigma }}{\delta \overline{c}}+\partial _\mu \frac{%
\delta \widehat{\Sigma }}{\delta \Omega _\mu }=0\;,  \label{antigh-eq}
\end{equation}

$\bullet \;$the linearly broken ghost Ward identity, typical of the Landau
gauge

\begin{equation}
\int d^4x\left( \frac{\delta \widehat{\Sigma }}{\delta c}+\left[ \overline{c}%
,\frac{\delta \widehat{\Sigma }}{\delta b}\right] \right) =\Delta _c^{cl}\;,
\label{gh-w-id}
\end{equation}
with $\Delta _c^{cl}\;$a linear classical breaking 
\begin{equation}
\Delta _c^{cl}=\int d^4x\left( \;[c,L]-[A,\Omega ]-[\phi ,D]+[\psi ,\xi ]-[%
\overline{\phi },\rho ]+[\eta ,\tau ]+\frac 12[\chi ,B]\right) \;.
\label{gh-lin-break}
\end{equation}
As usual, commuting the ghost equation $\;\left( \ref{gh-w-id}\right) \;$%
with the Slavnov-Taylor identity $\left( \ref{n-tym-s-t}\right) \;$one
obtains the Ward identity for the rigid gauge invariance \textit{\cite{book}}%
, expressing the fact that all the fields and antifields belong to the
adjoint representation of the gauge group.

\subsection{Classical approximation: the reduced action}

Following the standard procedure, let us introduce, for further use, the so
called reduced action $\widetilde{\mathcal{S}}\;$\textit{\cite{book}}
defined through the gauge fixing condition $\left( \ref{landau-g-f}\right) $
as

\begin{equation}
\widehat{\Sigma }=\widetilde{\mathcal{S}}\;+\;tr\int d^4x\;b\partial A\;,
\label{red-act}
\end{equation}
so that $\widetilde{\mathcal{S}}$\ is independent from the lagrangian
multiplier $b$. Moreover, from the antighost equation $\left( \ref{antigh-eq}%
\right) \;$it follows that $\widetilde{\mathcal{S}}\;$depends from the
antighost $\overline{c}\;$only through the combination$\footnote{%
The variable $\gamma_{\mu}$ is also called shifted antifield.}$ $\gamma ^\mu 
$

\begin{equation}
\gamma _\mu =\Omega _\mu +\partial _\mu \overline{c}\;.  \label{g-comb}
\end{equation}
Therefore, for $\widetilde{\mathcal{S}}\;$we have

\begin{eqnarray}
\widetilde{\mathcal{S}} &=&\frac 1{g^2}tr\displaystyle\int d^4x\;\left(
\frac 12F^{+}F^{+}-\chi ^{\mu \nu }(D_\mu \psi _\nu -D_\nu \psi _\mu
)^{+}+\eta D_\mu \psi ^\mu -\frac 12\overline{\phi }D_\mu D^\mu \phi
\;\right.  \nonumber \\
&&\;\;\;\;\;\;\;\;\left. \;+\frac 12\overline{\phi }\left\{ \psi ^\mu ,\psi
_\mu \right\} -\frac 12\phi \left\{ \chi ^{\mu \nu },\chi _{\mu \nu
}\right\} \;-\frac 18\left[ \phi ,\eta \right] \eta -\frac 1{32}\left[ \phi ,%
\overline{\phi }\right] ^2\right) \;  \nonumber \\
&&+tr\displaystyle\int d^4x\;\left( L(c^2-\omega ^2\phi -\omega \varepsilon
^\mu A_\mu +\frac{\varepsilon ^2}{16}\overline{\phi })+D(\left[ c,\phi
\right] -\varepsilon ^\mu \psi _\mu )\;\right.  \nonumber \\
&&\;\;\;\;\;\;\;\;\;+\gamma ^\mu (-D_\mu c+\omega \psi _\mu +\frac{%
\varepsilon ^\nu }2\chi _{\nu \mu }+\frac{\varepsilon _\mu }8\eta )\;+\;\rho
(\left[ c,\overline{\phi }\right] +2\omega \eta )\;  \nonumber \\
&&\;\;\;\;\;\;\;\;\;+\xi ^\mu (\left\{ c,\psi _\mu \right\} -\omega D_\mu
\phi +\varepsilon ^\nu F_{\nu \mu }-\frac{\varepsilon ^\nu }2F_{\nu \mu
}^{+}-\frac{\varepsilon _\mu }{16}[\phi ,\overline{\phi }])  \nonumber \\
&&\;\;\;\;\;\;\;\;\;+\tau (\left\{ c,\eta \right\} +\frac \omega 2[\phi ,%
\overline{\phi }]+\frac{\varepsilon ^\mu }2D_\mu \overline{\phi })\; 
\nonumber \\
&&\;\;\;\;\;\;\;\;\;\left. +\frac 12B^{\sigma \tau }(\left\{ c,\chi _{\sigma
\tau }\right\} +\omega F_{\sigma \tau }^{+}+\frac 18(\varepsilon ^\mu
\varepsilon _{\mu \sigma \tau \nu }+\varepsilon _\sigma g_{\nu \tau
}-\varepsilon _\tau g_{\nu \sigma })D^\nu \overline{\phi }\;)\right) 
\nonumber \\
&&+tr\displaystyle\int d^4x\left( \frac{g^2}8\omega ^2B^{\mu \nu }B_{\mu \nu
}-\frac{g^2}4\omega B^{\mu \nu }\varepsilon _\mu \xi _\nu -\frac{g^2}{32}%
\varepsilon ^\mu \varepsilon ^\nu \xi _\mu \xi _\nu +\frac{g^2}{32}%
\varepsilon ^2\xi ^2\right) \;.  \nonumber  \label{s-red-exp} \\
&&  \label{s-red-exp}
\end{eqnarray}
Accordingly, for the Slavnov-Taylor identity $\left( \ref{n-tym-s-t}\right) $
we get

\begin{equation}
\mathcal{S}(\widetilde{\mathcal{S}})\;\mathcal{=\;}\omega \varepsilon ^\mu 
\widetilde{\Delta }_\mu ^{cl}\;\;,  \label{red-s-t}
\end{equation}
where $\mathcal{S}(\widetilde{\mathcal{S}})$ denotes now the homogeneous
operator

\begin{eqnarray}
\mathcal{S}(\widetilde{\mathcal{S}}) &=&tr\int d^4x\left( \frac{\delta 
\widetilde{\mathcal{S}}}{\delta A^\mu }\frac{\delta \widetilde{\mathcal{S}}}{%
\delta \gamma _\mu }+\frac{\delta \widetilde{\mathcal{S}}}{\delta \xi ^\mu }%
\frac{\delta \widetilde{\mathcal{S}}}{\delta \psi _\mu }+\frac{\delta 
\widetilde{\mathcal{S}}}{\delta L}\frac{\delta \widetilde{\mathcal{S}}}{%
\delta c}+\frac{\delta \widetilde{\mathcal{S}}}{\delta D}\frac{\delta 
\widetilde{\mathcal{S}}}{\delta \phi }\;\right.  \nonumber
\label{red-s-t-op} \\
&&\;\;\;\;\;\;\;\;\;\;\;\;\;\;\left. +\frac{\delta \widetilde{\mathcal{S}}}{%
\delta \rho }\frac{\delta \widetilde{\mathcal{S}}}{\delta \overline{\phi }}%
\;\;+\frac{\delta \widetilde{\mathcal{S}}}{\delta \tau }\frac{\delta 
\widetilde{\mathcal{S}}}{\delta \eta }+\frac 12\frac{\delta \widetilde{%
\mathcal{S}}}{\delta B^{\mu \nu }}\frac{\delta \widetilde{\mathcal{S}}}{%
\delta \chi _{\mu \nu }}\right) \;\;,  \nonumber  \label{n-tym-st-op} \\
&&  \label{n-tym-st-op}
\end{eqnarray}
and $\widetilde{\Delta }_\mu ^{cl}\;$is given by the expression $\left( \ref
{v-break}\right) \;$with $\Omega _\mu \;$replaced by $\gamma _\mu $, \textit{%
i.e.}

\begin{eqnarray}
\widetilde{\Delta }_\mu ^{cl} &=&tr\int d^4x(\;L\partial _\mu c-D\partial
_\mu \phi -\gamma ^\nu \partial _\mu A_\nu +\xi ^\nu \partial _\mu \psi _\nu
\;  \nonumber \\
&&\;\;\;\;\;\;\;\;\;\;\;\;\;\;\;-\rho \partial _\mu \overline{\phi }+\tau
\partial _\mu \eta +\frac 12B^{\nu \sigma }\partial _\mu \chi _{\nu \sigma
}\;)\;.  \label{red-break}
\end{eqnarray}
For the linearized Slavnov-Taylor operator $\mathcal{B}_{\widetilde{\mathcal{%
S}}}$ we obtain now

\begin{eqnarray}
\mathcal{B}_{\widetilde{\mathcal{S}}} &=&tr\int d^4x\left( \frac{\delta 
\widetilde{\mathcal{S}}}{\delta A^\mu }\frac \delta {\delta \gamma _\mu }+%
\frac{\delta \widetilde{\mathcal{S}}}{\delta \gamma _\mu }\frac \delta
{\delta A^\mu }+\frac{\delta \widetilde{\mathcal{S}}}{\delta \psi _\mu }%
\frac \delta {\delta \xi ^\mu }+\frac{\delta \widetilde{\mathcal{S}}}{\delta
\xi ^\mu }\frac \delta {\delta \psi _\mu }+\frac{\delta \widetilde{\mathcal{S%
}}}{\delta L}\frac \delta {\delta c}\right.  \nonumber  \label{red-lin-op} \\
&&\;\;\;\;\;\;\;\;\;+\frac{\delta \widetilde{\mathcal{S}}}{\delta c}\frac
\delta {\delta L}+\frac{\delta \widetilde{\mathcal{S}}}{\delta \phi }\frac
\delta {\delta D}+\frac{\delta \widetilde{\mathcal{S}}}{\delta D}\frac
\delta {\delta \phi }+\frac{\delta \widetilde{\mathcal{S}}}{\delta \overline{%
\phi }}\frac \delta {\delta \rho }+\frac{\delta \widetilde{\mathcal{S}}}{%
\delta \rho }\frac \delta {\delta \overline{\phi }}+\frac{\delta \widetilde{%
\mathcal{S}}}{\delta \eta }\frac \delta {\delta \tau }  \nonumber \\
&&\;\;\;\;\;\;\;\;\left. \;+\frac{\delta \widetilde{\mathcal{S}}}{\delta
\tau }\frac \delta {\delta \eta }+\frac 12\frac{\delta \widetilde{\mathcal{S}%
}}{\delta \chi _{\mu \nu }}\frac \delta {\delta B^{\mu \nu }}+\frac 12\frac{%
\delta \widetilde{\mathcal{S}}}{\delta B^{\mu \nu }}\frac \delta {\delta
\chi _{\mu \nu }}\right) \;,  \nonumber \\
&&\;\;\;\;\;\;\;\;\;\;\;\;  \label{n-tym-lin-op}
\end{eqnarray}
and

\begin{equation}
\mathcal{B}_{\widetilde{\mathcal{S}}}\mathcal{B}_{\widetilde{\mathcal{S}}%
}=\omega \varepsilon ^\mu \mathcal{P}_\mu \;.  \label{red-lin-nil}
\end{equation}
The usefulness of working with the reduced action relies on the fact that $%
\widetilde{\mathcal{S}}\;$depends only on those variables which are really
relevant for the quantum analysis. It is apparent, for instance, that the
Landau gauge fixing condition $\left( \ref{landau-g-f}\right) \;$can be
regarded indeed as a linearly broken local Ward identity, implying thus that
the Lagrangian multiplier $b\;$cannot appear in the expression of the
invariant counterterms and of the possible anomalies. Of course, the same
holds for the antighost $\overline{c}\;$which, due to the equation\footnote{%
Both the Landau gauge-fixing condition and the antighost equation can be
proven to be renormalizable \textit{\cite{book}}.} $\left( \ref{antigh-eq}%
\right) $, can enter only through the combination $\gamma _\mu $.

Let us also recall, finally, the quantum numbers of all the fields and
antifields entering the expression of the reduced action $\left( \ref
{s-red-exp}\right) $.

\begin{equation}
\stackrel{Quantum\;numbers}{\stackunder{Table\;6.\;Fields}{
\begin{tabular}{|c|c|c|c|c|c|c|}
\hline
& $A_\mu $ & $\chi _{\mu \nu }$ & $\psi _\mu $ & $\eta $ & $\phi $ & $%
\overline{\phi }$ \\ \hline
$\dim \mathrm{.}$ & $1$ & $3/2$ & $3/2$ & $3/2$ & $1$ & $1$ \\ \hline
$\mathcal{R}-\mathrm{ch}\arg \mathrm{e}$ & $0$ & $-1$ & $1$ & $-1$ & $2$ & $%
-2$ \\ \hline
$\mathrm{gh-number}$ & $0$ & $0$ & $0$ & $0$ & $0$ & $0$ \\ \hline
$\mathrm{nature}$ & $comm.$ & $ant.$ & $ant.$ & $ant.$ & $comm.$ & $comm.$
\\ \hline
\end{tabular}
}}  \label{fields-table}
\end{equation}

\begin{equation}
\stackunder{Table\;7.\;Ghosts}{\stackrel{Quantum\;numbers}{
\begin{tabular}{|c|c|c|c|}
\hline
& $c$ & $\omega $ & $\varepsilon _\mu $ \\ \hline
$\dim .$ & $0$ & $-1/2$ & $-1/2$ \\ \hline
$\mathcal{R}-\mathrm{ch}\arg \mathrm{e}$ & $0$ & $-1$ & $1$ \\ \hline
$\mathrm{gh-number}$ & $1$ & $1$ & $1$ \\ \hline
$\mathrm{nature}$ & $ant.$ & $comm.$ & $comm.$ \\ \hline
\end{tabular}
}}  \label{ghosts-table}
\end{equation}

\begin{equation}
\stackrel{Quantum\;numbers}{\stackunder{Table\;8.\;Antifileds}{
\begin{tabular}{|c|c|c|c|c|c|c|c|}
\hline
& $L$ & $D$ & $\gamma ^\mu $ & $\xi ^\mu $ & $\rho $ & $\tau $ & $B^{\mu \nu
}$ \\ \hline
$\dim .$ & $4$ & $3$ & $3$ & $5/2$ & $3$ & $5/2$ & $5/2$ \\ \hline
$\mathcal{R}-\mathrm{ch}\arg \mathrm{e}$ & $0$ & $-2$ & $0$ & $-1$ & $2$ & $%
1 $ & $1$ \\ \hline
$\mathrm{gh-number}$ & $-2$ & $-1$ & $-1$ & $-1$ & $-1$ & $-1$ & $-1$ \\ 
\hline
$\mathrm{nature}$ & $comm.$ & $ant.$ & $ant.$ & $comm.$ & $ant.$ & $comm.$ & 
$comm.$ \\ \hline
\end{tabular}
}}  \label{antifields-table}
\end{equation}

\newpage\ 

\section{Renormalization of topological Yang-Mills}

\subsection{Cohomology of the linearized Slavnov-Taylor operator $\mathcal{B}%
_{\widetilde{\mathcal{S}}}$}

We are now ready to discuss the renormalization of the TYM. As already
underlined in Sect.2, the first task is that of characterizing the
cohomology classes of the linearized Slavnov-Taylor operator which turn out
to be relevant for the anomalies and the invariant counterterms. Let us
recall that both anomalies and invariant counterterms are integrated local
polynomials $\Delta ^G\;$in the fields $(A,\psi ,\chi ,\eta ,\phi ,\overline{%
\phi },c)$, in the antifields $(L,D,\gamma ,\xi ,\rho ,\tau ,B)$,$\;$and in
the global ghosts $(\omega ,\varepsilon )$,$\;$with dimension four,
vanishing $\mathcal{R}$-charge and ghost number $G\;$respectively one and
zero. In addition, they are constrained by the consistency condition

\begin{equation}
\mathcal{B}_{\widetilde{\mathcal{S}}}\Delta
^G=0\;,\;\;\;\;\;\;\;\;\;\;\;G=0,1\;\;,  \label{int-cons-cond}
\end{equation}
$\mathcal{B}_{\widetilde{\mathcal{S}}}\;$being the linearized Slavnov-Taylor
operator of eq.$\left( \ref{n-tym-lin-op}\right) $. From the relation $%
\left( \ref{red-lin-nil}\right) $, \textit{i.e.}

\begin{equation}
\mathcal{B}_{\widetilde{\mathcal{S}}}\mathcal{B}_{\widetilde{\mathcal{S}}%
}=\omega \varepsilon ^\mu \mathcal{P}_\mu \;\;,  \label{sec-red-lin-nil}
\end{equation}
one sees that $\mathcal{B}_{\widetilde{\mathcal{S}}}\;$is in fact a
nilpotent operator when acting on the space of the integrated functionals
which are invariant under space-time translations. It follows therefore that
the relevant solutions of the eq.$\left( \ref{int-cons-cond}\right) \;$%
identify nontrivial elements of the integrated cohomology of $\mathcal{B}_{%
\widetilde{\mathcal{S}}}$. In order to characterize the integrated
cohomology of $\mathcal{B}_{\widetilde{\mathcal{S}}}$ we introduce the
operator

\begin{equation}
\mathcal{N}_\varepsilon =\varepsilon ^\mu \frac \partial {\partial
\varepsilon ^\mu }\;,  \label{count-op}
\end{equation}
which counts the number of global ghosts $\varepsilon ^\mu \;$contained in a
given field polynomial. Accordingly, the functional operator $\mathcal{B}_{%
\widetilde{\mathcal{S}}}\;$displays the following $\varepsilon $-expansion

\begin{equation}
\mathcal{B}_{\widetilde{\mathcal{S}}}=b_{_{\widetilde{\mathcal{S}}%
}}+\varepsilon ^\mu \mathcal{W}_\mu +\frac 12\varepsilon ^\mu \varepsilon
^\nu \mathcal{W}_{\mu \nu }\;,  \label{lin-dec}
\end{equation}
where, from eq.$\left( \ref{sec-red-lin-nil}\right) \;$the operators $b_{_{%
\widetilde{\mathcal{S}}}},\;\mathcal{W}_\mu ,\;\mathcal{W}_{\mu \nu }\;$are
easily seen to obey the following algebraic relations

\begin{equation}
b_{_{\widetilde{\mathcal{S}}}}b_{_{\widetilde{\mathcal{S}}}}=0\;,
\label{b-exact-nilp}
\end{equation}

\begin{equation}
\left\{ b_{_{\widetilde{\mathcal{S}}}},\mathcal{W}_\mu \right\} =\omega 
\mathcal{P}_\mu \;,  \label{b-W-dec}
\end{equation}
and

\begin{eqnarray}
\left\{ \mathcal{W}_\mu ,\mathcal{W}_\nu \right\} +\left\{ b_{_{\widetilde{%
\mathcal{S}}}},\mathcal{W}_{\mu \nu }\right\} &=&0\;,  \label{W-alg} \\
\left\{ \mathcal{W}_\mu ,\mathcal{W}_{\nu \rho }\right\} +\left\{ \mathcal{W}%
_\nu ,\mathcal{W}_{\rho \mu }\right\} +\left\{ \mathcal{W}_\rho ,\mathcal{W}%
_{\mu \nu }\right\} &=&0\;,  \nonumber \\
\left\{ \mathcal{W}_{\mu \nu },\mathcal{W}_{\rho \sigma }\right\} +\left\{ 
\mathcal{W}_{\mu \rho },\mathcal{W}_{\nu \sigma }\right\} +\left\{ \mathcal{W%
}_{\mu \sigma },\mathcal{W}_{\nu \rho }\right\} &=&0\;.  \nonumber
\end{eqnarray}
with

\[
\stackrel{Quantum\;numbers}{\stackunder{Table\;9.}{
\begin{tabular}{|c|c|c|c|}
\hline
& $b_{_{\widetilde{\mathcal{S}}}}$ & $\mathcal{W}_\mu $ & $\mathcal{W}_{\mu
\nu }$ \\ \hline
$\dim .$ & $0$ & $1/2$ & $1$ \\ \hline
$\mathcal{R}-\mathrm{ch}\arg \mathrm{e}$ & $0$ & $-1$ & $-2$ \\ \hline
$\mathrm{gh-number}$ & $1$ & $0$ & $-1$ \\ \hline
$\mathrm{nature}$ & $ant.$ & $ant.$ & $ant.$ \\ \hline
\end{tabular}
\;}} 
\]
In particular, from the eqs.$\left( \ref{b-exact-nilp}\right) ,\;\left( \ref
{b-W-dec}\right) \;$we observe that the operator $b_{_{\widetilde{\mathcal{S}%
}}}\;$is strictly nilpotent and that the vector operator $\mathcal{W}_\mu \;$%
allows to decompose the space-time translations $\mathcal{P}_\mu \;$as a $%
b_{_{\widetilde{\mathcal{S}}}}-$anticommutator, providing thus an off-shell
realization of the algebra $\left( \ref{wz-talg-1}\right) $. Furthermore, we
shall check that the decomposition formula $\left( \ref{b-W-dec}\right) \;$%
will be of great usefulness in order to obtain the nontrivial expression of
the invariant counterterms of the complete operator $\mathcal{B}_{\widetilde{%
\mathcal{S}}}$. Let us also give, for further use, the explicit form of the
operators $b_{_{\widetilde{\mathcal{S}}}},\;\mathcal{W}_\mu ,\;\mathcal{W}%
_{\mu \nu }$, namely

\begin{eqnarray}
b_{_{\widetilde{\mathcal{S}}}}A_\mu &=&-D_\mu c+\omega \psi _\mu \;,
\label{b-tr-fields} \\
b_{_{\widetilde{\mathcal{S}}}}\psi _\mu &=&\left\{ c,\psi _\mu \right\}
-\omega D_\mu \phi \;,  \nonumber \\
b_{_{\widetilde{\mathcal{S}}}}c &=&c^2-\omega ^2\phi \;,  \nonumber \\
b_{_{\widetilde{\mathcal{S}}}}\phi &=&\left[ c,\phi \right] \;,  \nonumber \\
b_{_{\widetilde{\mathcal{S}}}}\overline{\phi } &=&\left[ c,\overline{\phi }%
\right] +2\omega \eta \;,  \nonumber \\
b_{_{\widetilde{\mathcal{S}}}}\eta &=&\left\{ c,\eta \right\} +\frac \omega
2\left[ \phi ,\overline{\phi }\right] \;,  \nonumber \\
b_{_{\widetilde{\mathcal{S}}}}\chi _{\sigma \tau } &=&\left\{ c,\chi
_{\sigma \tau }\right\} +\omega F_{\sigma \tau }^{+}+\frac{g^2}2\omega
^2B_{\sigma \tau }\;,  \nonumber
\end{eqnarray}
and

\begin{eqnarray}
b_{_{\widetilde{\mathcal{S}}}}\gamma _\mu &=&\frac 1{g^2}\left( 4D^\nu
F_{\mu \nu }+4\left\{ \psi ^\nu ,\chi _{\mu \nu }\right\} -\left\{ \psi _\mu
,\eta \right\} +\frac 12\left[ \overline{\phi },D_\mu \phi \right] -\frac
12\left[ D_\mu \overline{\phi },\phi \right] \right)  \nonumber
\label{b-tr-antifields} \\
&&-\omega \left[ \phi ,\xi _\mu \right] \;+2\omega D^\nu B_{\mu \nu
}+\left\{ c,\gamma _\mu \right\} \;,  \nonumber \\
b_{_{\widetilde{\mathcal{S}}}}\xi _\mu &=&\frac 1{g^2}\left( 4D^\nu \chi
_{\mu \nu }+D_\mu \eta -\left[ \overline{\phi },\psi _\mu \right] \right)
-\omega \gamma _\mu +\left[ c,\xi _\mu \right] \;,  \nonumber \\
b_{_{\widetilde{\mathcal{S}}}}L &=&\left[ c,L\right] -\left[ \phi ,D\right]
-D^\mu \gamma _\mu +\left[ \psi _\mu ,\xi ^\mu \right] -\left[ \overline{%
\phi },\rho \right] +\left[ \eta ,\tau \right] +\frac 12\left[ \chi ^{\sigma
\tau },B_{\sigma \tau }\right] \;,  \nonumber \\
b_{_{\widetilde{\mathcal{S}}}}D &=&\frac 1{g^2}\left( -\frac 12D^\mu D_\mu 
\overline{\phi }-\frac 12\left\{ \chi ^{\mu \nu },\chi _{\mu \nu }\right\}
-\frac 18\left\{ \eta ,\eta \right\} -\frac 1{16}\left[ \overline{\phi }%
,\left[ \phi ,\overline{\phi }\right] \right] \right)  \nonumber \\
&&+\omega D^\mu \xi _\mu -\omega ^2L+\frac \omega 2\left[ \overline{\phi }%
,\tau \right] \;+\left\{ \;c,D\right\} \;,\;  \nonumber \\
b_{_{\widetilde{\mathcal{S}}}}\rho &=&\frac 1{2g^2}\left( -D^\mu D_\mu \phi
+\left\{ \psi ^\mu ,\psi _\mu \right\} +\frac 18\left[ \phi ,\left[ \phi ,%
\overline{\phi }\right] \right] \right) -\frac \omega 2\left[ \phi ,\tau
\right] +\left\{ c,\rho \right\} \;,  \nonumber \\
b_{_{\widetilde{\mathcal{S}}}}\tau &=&\frac 1{g^2}\left( D^\mu \psi _\mu
+\frac 14\left[ \phi ,\eta \right] \right) -2\omega \rho +\left[ c,\tau
\right] \;,  \label{b-tr-antifields} \\
b_{_{\widetilde{\mathcal{S}}}}B_{\mu \nu } &=&\frac 1{g^2}\left( -2(D_\mu
\psi _\nu -D_\nu \psi _\mu )^{+}+2\left[ \phi ,\chi _{\mu \nu }\right]
\right) +\left[ c,B_{\mu \nu }\right] \;.  \nonumber
\end{eqnarray}
For the operator $\mathcal{W}_\mu \;$we get

\begin{eqnarray}
\mathcal{W}_\mu A_\nu &=&\frac 12\chi _{\mu \nu }+\frac 18g_{\mu \nu }\eta
\;,  \nonumber  \label{W-tr-fields} \\
\mathcal{W}_\mu \psi _\nu &=&F_{\mu \nu }-\frac 12F_{\mu \nu }^{+}-\frac
1{16}g_{\mu \nu }\left[ \phi ,\overline{\phi }\right] -\frac{g^2}4\omega
B_{\mu \nu }\;,  \nonumber \\
\mathcal{W}_\mu c &=&-\omega A_\mu \;,  \nonumber \\
\mathcal{W}_\mu \phi &=&-\psi _\mu \;,  \nonumber \\
\mathcal{W}_\mu \overline{\phi } &=&0\;,  \nonumber \\
\mathcal{W}_\mu \eta &=&\frac 12D_\mu \overline{\phi }\;,  \nonumber \\
\mathcal{W}_\mu \chi _{\sigma \tau } &=&\frac 18\left( \varepsilon _{\mu
\sigma \tau \nu }D^\nu \overline{\phi }+g_{\mu \sigma }D_\tau \overline{\phi 
}-g_{\mu \tau }D_\sigma \overline{\phi }\right)  \nonumber \\
&&-\frac{g^2}8\omega \left( \varepsilon _{\mu \sigma \tau \nu }\xi ^\nu
+\eta _{\mu \sigma }\xi _\tau -\eta _{\mu \tau }\xi _\sigma \right) \;,
\label{W-tr-fields}
\end{eqnarray}
and

\begin{eqnarray}
\mathcal{W}_\mu \gamma _\sigma &=&\left( -\omega g_{\mu \sigma }L-\frac
12D_\mu \xi _\sigma +\frac 12g_{\mu \sigma }D^\tau \xi _\tau +\frac 12g_{\mu
\sigma }\left[ \overline{\phi },\tau \right] \right.  \nonumber \\
&&\left. -\frac 14\left[ B_{\mu \sigma ,}\overline{\phi }\right] -\frac
12\varepsilon _{\mu \sigma \nu \rho }D^\nu \xi ^\rho \;,\right)  \nonumber \\
\mathcal{W}_\mu \xi _\sigma &=&g_{\mu \sigma }D\;,  \nonumber \\
\mathcal{W}_\mu L &=&0\;,  \nonumber \\
\mathcal{W}_\mu D &=&-\frac 1{16}\left[ \overline{\phi },\xi _\mu \right] \;,
\nonumber \\
\mathcal{W}_\mu \rho &=&\frac 1{16}\left[ \phi ,\xi _\mu \right] -\frac
12D_\mu \tau -\frac 14D^\nu B_{\mu \nu }\;,  \nonumber \\
\mathcal{W}_\mu \tau &=&-\frac 18\gamma _\mu \;,  \label{W-tr-antifields} \\
\mathcal{W}_\mu B_{\sigma \tau } &=&-\frac 14\left( \varepsilon _{\mu \sigma
\tau \nu }\gamma ^\nu +g_{\mu \sigma }\gamma _\tau -g_{\mu \tau }\gamma
_\sigma \right) \;.  \nonumber
\end{eqnarray}
Finally, for $\mathcal{W}_{\mu \nu }\;$one obtains

\begin{eqnarray}
\mathcal{W}_{\mu \nu }A_\sigma &=&\mathcal{W}_{\mu \nu }\phi =\mathcal{W}%
_{\mu \nu }\overline{\phi }=\mathcal{W}_{\mu \nu }\eta =\mathcal{W}_{\mu \nu
}\chi _{\sigma \tau }=0\;,  \nonumber  \label{WW-tr-fields} \\
\mathcal{W}_{\mu \nu }\psi _\sigma &=&\frac{g^2}{16}\left( 2g_{\mu \nu }\xi
_\sigma -g_{\mu \sigma }\xi _\nu -g_{\nu \sigma }\xi _\mu \right) \;, 
\nonumber \\
\mathcal{W}_{\mu \nu }c &=&\frac 18g_{\mu \nu }\overline{\phi }\;,
\label{WW-tr-fields}
\end{eqnarray}
and, for the antifields

\begin{eqnarray}
\mathcal{W}_{\mu \nu }\gamma _\sigma &=&\mathcal{W}_{\mu \nu }\xi _\sigma =%
\mathcal{W}_{\mu \nu }L=\mathcal{W}_{\mu \nu }D=\mathcal{W}_{\mu \nu }\tau =%
\mathcal{W}_{\mu \nu }B_{\sigma \tau }=0\;,  \nonumber \\
\mathcal{W}_{\mu \nu }\rho &=&\frac 18g_{\mu \nu }L\;.
\label{WW-tr-antifields}
\end{eqnarray}

According to the general results on cohomology given in the Subsect. 2.3,
the integrated cohomology of \ $\mathcal{B}_{\widetilde{\mathcal{S}}}\;$is
isomorphic to a subspace of the integrated cohomology of \ $b_{_{\widetilde{%
\mathcal{S}}}}$. Let us therefore first focus on the operator $b_{_{%
\widetilde{\mathcal{S}}}}$. We have now to mention that one of the
advantages of having decomposed the operator $\mathcal{B}_{\widetilde{%
\mathcal{S}}}\;$according to the counting operator $\left( \ref{count-op}%
\right) $\ is due to the fact that the (nonintegrated) cohomology classes of 
$b_{_{\widetilde{\mathcal{S}}}}\;$have been already identified by the
authors of ref. \textit{\cite{dmpw}}, who computed in fact the cohomology of 
$b_{_{\widetilde{\mathcal{S}}}}\;$in terms of the so called \textit{%
invariant }or \textit{constrained }cohomology, the name invariant cohomology
standing for the computation of the cohomology in the space of the gauge
invariant polynomials. Their result can be immediately adapted to our
present case, being stated as follows:

\begin{itemize}
\item  \textit{The cohomology classes of the operator }$b_{_{\widetilde{%
\mathcal{S}}}}$\ \textit{in the space of the nonintegrated local polynomials
in the fields and antifields which are analytic in the global ghosts are
given by invariant polynomials in the undifferentiated field }$\phi \;$%
\textit{built up with monomials }$\mathcal{P}_n(\phi )$\ \textit{of the type 
} 
\begin{equation}
\mathcal{P}_n(\phi )=tr\left( \frac{\phi ^n}n\right) \;,\;\;\;\;\;\;n\geq
2\;.  \label{b-cohomology}
\end{equation}

\begin{remark}
Although the formal proof of the above result can be found in the original
work \textit{\cite{dmpw}}, let us present here a very simple and intuitive
argument for a better understanding of eq.$\left( \ref{b-cohomology}\right)
. $ Following \textit{\cite{dmpw}} we further decompose the operator $b_{_{%
\widetilde{\mathcal{S}}}}$ according to the counting operator $\mathcal{N}%
=\omega \partial /\partial \omega $, i.e. 
\begin{equation}
b_{_{\widetilde{\mathcal{S}}}}=b_{_{\widetilde{\mathcal{S}}}}^0\;+\omega
b_{_{\widetilde{\mathcal{S}}}}^1\;+\omega ^2b_{_{\widetilde{\mathcal{S}}%
}}^2\;.  \label{b-decomp}
\end{equation}
From eqs.$\left( \ref{b-tr-fields}\right) ,\left( \ref{b-tr-antifields}%
\right) $ it is apparent to see that the first term $b_{_{\widetilde{%
\mathcal{S}}}}^0\;$of the decomposition $\left( \ref{b-decomp}\right) $
picks up the part of the operator $b_{_{\widetilde{\mathcal{S}}}}$
corresponding to the pure gauge transformations, while the second and the
third term $(b_{_{\widetilde{\mathcal{S}}}}^1,b_{_{\widetilde{\mathcal{S}}%
}}^2)\;$have basically the effect of a shift transformation. This is
particularly evident if one looks at the first set of transformations $%
\left( \ref{b-tr-fields}\right) \;$concerning only the fields $(A,\psi ,\chi
,\eta ,\phi ,\overline{\phi },c).$ It should also be remarked that among the
fields $(A,\psi ,\chi ,\eta ,\phi ,\overline{\phi },c)\;$the scalar
component $\phi $ is the only field whose transformation does not contain
the global ghost $\omega $, so that the action of the operator $b_{_{%
\widetilde{\mathcal{S}}}}\;$on $\phi \;$reduces to a simple pure gauge
transformation. Recalling now that the cohomology of $b_{_{\widetilde{%
\mathcal{S}}}}\;$is, in turn, isomorphic to a subspace of the cohomology of $%
b_{_{\widetilde{\mathcal{S}}}}^0$, we can easily understand that the
characterization of the cohomology of $b_{_{\widetilde{\mathcal{S}}}}\;$can
be reduced in fact to a computation of the so called invariant cohomology,
i.e. to the computation of the cohomology in the space of the gauge
invariant polynomials which are left unchanged by the shift transformations
corresponding to $(b_{_{\widetilde{\mathcal{S}}}}^1,b_{_{\widetilde{\mathcal{%
S}}}}^2)$. It follows therefore that the only invariants which survive are
exactely the polynomials in the undifferentiated field $\phi $, justifying
thus the result $\left( \ref{b-cohomology}\right) $. We also remark that the
polynomials $\mathcal{P}_n(\phi )\;$can be eventually multiplied by
appropriate powers in the constant ghosts $(\varepsilon _\mu ,\omega )$ in
order to obtain invariants with the right $\mathcal{R}$-charge, ghost number
and dimension. Finally let us underline that, due to the commuting nature of
the field $\phi $, the expression $tr(\phi ^n)\;$(for $n$ sufficiently
large) is related to higher order invariant Casimir tensors whose existence
relies on the choice of the gauge group G.
\end{remark}
\end{itemize}

\subsection{Analyticity in the constant ghosts and triviality of the
cohomology of $b_{_{\widetilde{\mathcal{S}}}}$}

Before analysing the consequences which follow from the result $\left( \ref
{b-cohomology}\right) \;$on the cohomology of the complete operator $%
\mathcal{B}_{\widetilde{\mathcal{S}}}$, let us discuss here the important
issue of the analyticity in the constant global ghosts. In fact, as
repeatedly mentioned in the previous Sections, the requirement of
analyticity in the ghosts $(\varepsilon _\mu ,\omega )$, stemming from pure
perturbative considerations, is one of the most important ingredient in
order to interpret the TYM in terms of a standard field theory which can be
characterized by a nonvanishing BRST cohomology. In particular, we have
already emphasized that the nonemptiness of the BRST cohomology relies
exactly on the analyticity requirement. For a better understanding of this
point, let us consider in detail the case of the simplest invariant
polynomial $\mathcal{P}_2(\phi )\;$

\begin{equation}
\mathcal{P}_2(\phi )\;=\frac 12tr\phi ^2\;.  \label{p2-inv-pol}
\end{equation}
It is an almost trivial exercise to show that $tr\phi ^2\;$can be expressed
indeed as a pure $b_{_{\widetilde{\mathcal{S}}}}-$variation, namely

\begin{equation}
tr\phi ^2=b_{_{\widetilde{\mathcal{S}}}}tr\left( -\frac 1{\omega ^2}c\phi
+\frac 1{3\omega ^4}c^3\right) \;.  \label{triv-p2}
\end{equation}
This formula illustrates in a very clear way the relevance of the
analyticity requirement. It is apparent from the eq.$\left( \ref{triv-p2}%
\right) \;$that the price to be payed in order to write $tr\phi ^2\;$as a
pure $b_{_{\widetilde{\mathcal{S}}}}-$variation is in fact the loss of
analyticity in the global ghost $\omega $.

In other words, as long as one works in a functional space whose elements
are power series in the global ghosts, the cohomology of $b_{_{\widetilde{%
\mathcal{S}}}}\;$is not empty. As we shall see later on, this will imply
that also the cohomology of $\mathcal{B}_{\widetilde{\mathcal{S}}}$ will be
nontrivial, meaning that TYM can be regarded as a standard supersymmetric
gauge theory of the Yang-Mills type. On the other hand, if the analyticity
requirement is relaxed, the cohomology of $b_{_{\widetilde{\mathcal{S}}}}$,
and therefore that of the complete operator $\mathcal{B}_{\widetilde{%
\mathcal{S}}}$, becomes trivial, leading thus to the cohomological
interpretation of Baulieu-Singer \textit{\cite{bs}} and Labastida-Pernici 
\textit{\cite{lp}}. We see therefore that the analyticity requirement in the
global ghosts is the property which intertwines the two possible
interpretation of TYM. One goes from the standard field theory point of view
to the cohomological one by simply setting $\omega =1$, which of course
implies that analyticity is lost. In addition, it is rather simple to
convince oneself that setting $\omega =1$ has the meaning of identifying the 
$\mathcal{R}$-charge with the ghost number, so that the fields $(\chi ,\psi
,\eta ,\phi ,\overline{\phi })\;$aquire a nonvanishing ghost number given
respectively by $(-1,1,-1,2,-2)$. They correspond now to the so called
topological ghosts of the cohomological interpretation.

\begin{remark}
It is also interesting to point out that there is a deep relationship
between the analyticity in the global ghosts and the so called equivariant
cohomology proposed by R. Stora et al. \textit{\cite{stora1,stora2}} in
order to recover the Witten's observables in the case in which TYM\ is
considered as a cohomological theory with vanishing BRST cohomology. Roughly
speaking, the equivariant cohomology can be defined as the restriction of
the BRST\ cohomology to the space of the gauge invariant polynomials which
cannot be written as the BRST variation of local quantities which are
independent from the Faddeev-Popov ghost $c$. In other words, a gauge
invariant cocycle $\vartheta \;$is called nontrivial in the equivariant
cohomology if $\vartheta $ cannot be written as the $b_{_{\widetilde{%
\mathcal{S}}}}$-variation of a local polynomial $\widetilde{\vartheta }\;$%
which is independent from the ghost $c$, i.e. if 
\begin{equation}
\vartheta =b_{_{\widetilde{\mathcal{S}}}}\widetilde{\vartheta }\;,
\label{eqiuv-cocycle}
\end{equation}
with $\widetilde{\vartheta }\;$containing necessarily $c$, then $\vartheta $
identifies a nontrivial element of the equivariant cohomology\footnote{%
See ref.\textit{\cite{stora2}} for a geometrical construction of the
equivariant cohomology classes.}. Considering now the polynomial $tr\phi ^2$%
, we see that it yields a nontrivial equivariant cocycle in the
cohomological interpretation $(i.e.\;\omega =1)$, due to the unavoidable
presence of the Faddeev-Popov ghost $c$ on the right hand side of eq.$\left( 
\ref{triv-p2}\right) $. Moreover, keeping the standard point of view $%
(i.e.\;\omega \neq 1)$, it is apparent that in order to write $tr\phi ^2\;$%
as a trivial cocycle use has to be done of the effective variable $c/\omega $%
, meaning that the loss of anlyticity is accompanied by the presence of the
Faddeev-Popov ghost $c$. This shows that, in the case of the invariant
polynomials $\mathcal{P}_n(\phi ),\;$the analyticity requirement and the
equivariant cohomology identify indeed the same class of invariants \textit{%
\cite{bc}}.
\end{remark}

\subsection{The integrated cohomology of $b_{_{\widetilde{\mathcal{S}}}}$
and the absence of anomalies.}

Having characterized the cohomology of $b_{_{\widetilde{\mathcal{S}}}}$, let
us now briefly analyse the integrated cohomology classes or, equivalently,
the local cocycles which belong to the cohomology of $b_{_{\widetilde{%
\mathcal{S}}}}$ modulo a total space-time derivative. Let us begin first
with the integrated cohomology classes which have the same quantum numbers
of the invariant counterterms, \textit{i.e.} dimension four and vanishing $%
\mathcal{R}-$charge and ghost number. Combining the result $\left( \ref
{b-cohomology}\right) $ with the technique of the descent equations\footnote{%
See Chapt.5 of ref.\textit{\cite{book}} for a self contained illustration of
the method.}, it turns out that the integrated cohomology classes of $b_{_{%
\widetilde{\mathcal{S}}}}$ with dimension four and vanishing ghost number
can be identified, modulo exact cocycles, with the following three elements,

\begin{equation}
\int d^4x\;tr(\phi ^4)\;,\;\;\;\;\;\;\;\;\int d^4x\;\left( tr\phi ^2\right)
^{2\;},  \label{disg-el}
\end{equation}
and

\begin{equation}
\varepsilon ^{\mu \nu \rho \tau }\mathcal{W}_\mu \mathcal{W}_\nu \mathcal{W}%
_\rho \mathcal{W}_\tau \int d^4x\;tr\left( \frac{\phi ^{2\;}}2\right) \;,
\label{nontr-el}
\end{equation}
with $\mathcal{W}_\mu $ given in eqs.$\left( \ref{lin-dec}\right) ,\left( 
\ref{W-tr-fields}\right) ,\left( \ref{W-tr-antifields}\right) $. The first
two terms in eq. $\left( \ref{disg-el}\right) $ possess $\mathcal{R}$-charge
8 and therefore have to be ruled out. The only term with the correct quantum
numbers (see Tables 6-8 of eqs.$\left( \ref{fields-table}\right) $-$\left( 
\ref{antifields-table}\right) $) is thus that of eq.$\left( \ref{nontr-el}%
\right) .$ The invariance of the term $\left( \ref{nontr-el}\right) $ under
the action of $b_{_{\widetilde{\mathcal{S}}}}$ easily follows from the
decomposition $\left( \ref{b-W-dec}\right) $, which implies that $b_{_{%
\widetilde{\mathcal{S}}}}$ and $\mathcal{W}_\mu $ can be regarded as
anticommuting operators when acting on the space of the integrated local
functionals. Its nontriviality is easily seen to be a consequence of the
nontriviality of $\left( tr\phi ^2\right) $, according to eq.$\left( \ref
{b-cohomology}\right) .$

\begin{remark}
It is worth to underline here that expressions of the type of eq.$\left( \ref
{nontr-el}\right) \;$are not a novelty, as they appear rather naturally in
the study of the BRST cohomology of the topological models. Their occurrence
lies precisely on the existence of a vector operator $\mathcal{W}_\mu \;$%
which allows to decompose the space-time derivatives as a BRST
anticommutator, as in the equation $\left( \ref{b-W-dec}\right) $. As shown
in \textit{\cite{st}}, the decomposition formula $\left( \ref{b-W-dec}%
\right) $\ turns out to be of great importance in order to obtain the
integrated cohomology classes from the nonintegrated ones. The operator $%
\mathcal{W}_\mu \;$plays in fact the role of a climbing operator which
allows to solve in a very straightforward and elegant way the descent
equations corresponding to the integrated cohomology. It is a simple
exercise, for instance, to verify that the following cocycles, respectively
a one, two and a three form, 
\begin{eqnarray}
&&\mathcal{W}_\mu \left( tr\frac{\phi ^{2\;}}2\right) dx^\mu \;,
\label{w-obs} \\
&&\mathcal{W}_\mu \mathcal{W}_\nu \left( tr\frac{\phi ^{2\;}}2\right) dx^\mu
\wedge dx^\nu \;,  \nonumber \\
&&\mathcal{W}_\mu \mathcal{W}_\nu \mathcal{W}_\rho \left( tr\frac{\phi ^{2\;}%
}2\right) dx^\mu \wedge dx^\nu \wedge dx^\rho \;,\;  \nonumber
\end{eqnarray}
belong indeed to the cohomology of $b_{_{\widetilde{\mathcal{S}}}}\;$modulo
a total derivative. The expression $\left( \ref{w-obs}\right) \;$are seen to
reproduce (modulo trivial terms) the Witten observables of the TYM theory 
\textit{\cite{tym}}, giving a more direct idea of the usefulness of the
operator $\mathcal{W}_\mu $.
\end{remark}

\vspace{5mm}

Turning now to the integrated cohomology classes in the sector of dimension
four, $\mathcal{R}$-charge zero and ghost number one, it can be proven that
the result $\left( \ref{b-cohomology}\right) $ implies that the only
integrated invariants which can be defined are those which are $b_{_{%
\widetilde{\mathcal{S}}}}$-exact, meaning that the integrated cohomology of $%
b_{_{\widetilde{\mathcal{S}}}}$ in the sector of the anomalies is empty.
Therefore the integrated cohomology of the complete operator $\mathcal{B}_{%
\widetilde{\mathcal{S}}}\;$turns out to be empty as well, so that the
classical Slavnov-Taylor identity $\left( \ref{n-tym-s-t}\right) $ can be
extended at the quantum level without anomalies. It is important to mention
that this result, already obtained by N. Maggiore \textit{\cite{magg}} in
the analysis of the $N=2$ untwisted gauge theories, means, in particular,
that there is no possible extension of the nonabelian Adler-Bardeen gauge
anomaly compatible with $N=2$ supersymmetry.

In summary, we have seen that the operator $b_{_{\widetilde{\mathcal{S}}}}$
has a nonvanishing integrated cohomology only in the sector of the invariant
counterterms, in which the unique nontrivial element is given by the
expression $\left( \ref{nontr-el}\right) $. We have now to recall that the
computation of the cohomology of $b_{_{\widetilde{\mathcal{S}}}}$ is only a
first step towards the characterization of the nontrivial classes of the
complete operator $\mathcal{B}_{\widetilde{\mathcal{S}}}$. Moreover, from
the previous results, we can infer that the cohomology of $\mathcal{B}_{%
\widetilde{\mathcal{S}}}$ in the sector of the invariant counterterms can
contain at most a unique element. The task of the next final section will be
that of providing the expression of the unique nontrivial invariant
counterterm of TYM.

\subsection{The invariant counterterm of TYM}

Let us face now the problem of the characterization of the most general
local invariant counterterm $\Delta ^{count}$ which can be freely added, to
each order of perturbation theory, to the vertex functional $\Gamma $ which
fulfils the quantum version of the Slavnov-Taylor identity\footnote{%
We remind here that, due to the absence of anomalies, the classical
Slavnov-Taylor identity $\left( \ref{n-tym-s-t}\right) $ can be always
extended at the quantum level.} $\left( \ref{n-tym-s-t}\right) $,

\begin{eqnarray}
\mathcal{S}(\Gamma )\; &=&\mathcal{\;}\omega \varepsilon ^\mu \Delta _\mu
^{cl}\;\;,  \label{quant-s-t} \\
\Gamma &=&\widehat{\Sigma }+O(\hbar )\;.\;\mathcal{\;}  \nonumber
\end{eqnarray}
We look then for an integrated local polynomial in the fields, antifields
and global ghosts with dimension four and vanishing $\mathcal{R}$-charge and
ghost number, which is a nontrivial solution of the consistency condition

\begin{equation}
\mathcal{B}_{\widetilde{\mathcal{S}}}\Delta ^{count}=0\;.
\label{inv-count-cons}
\end{equation}
In order to find a candidate for $\Delta ^{count}\;$we observe that the
classical breaking term in the right-hand side of eqs.$\left( \ref{n-tym-s-t}%
\right) $ and $\left( \ref{quant-s-t}\right) \;$does not depend from the
gauge coupling constant $g$ of TYM, according to eq.$\left( \ref{v-break}%
\right) $. Therefore, acting with $\partial /\partial g$ on both side of eq.$%
\left( \ref{red-s-t}\right) $ we get

\begin{equation}
\mathcal{B}_{\widetilde{\mathcal{S}}}\frac{\partial \widetilde{\mathcal{S}}}{%
\partial g}=0\;.  \label{nice-way}
\end{equation}
We see that $\partial \widetilde{\mathcal{S}}/\partial g$ yields a solution
of the consistency condition $\left( \ref{inv-count-cons}\right) $. Of
course, it remains to prove that $\partial \widetilde{\mathcal{S}}/\partial
g\;$identifies a nontrivial element of the cohomology of $\mathcal{B}_{%
\widetilde{\mathcal{S}}}$. In order to prove the nontriviality of $\partial 
\widetilde{\mathcal{S}}/\partial g\;$we proceed, following a well known
standard cohomology argument, by assuming the converse, \textit{i.e. }that $%
\partial \widetilde{\mathcal{S}}/\partial g\;$can be written as an exact
cocycle:

\begin{equation}
\frac{\partial \widetilde{\mathcal{S}}}{\partial g}=\mathcal{B}_{\widetilde{%
\mathcal{S}}}\Xi ^{-1}\;,  \label{conv}
\end{equation}
for some integrated local polynomial $\Xi ^{-1}$ in the fields, antifields
and constant ghosts with negative ghost number. Decomposing now the equation 
$\left( \ref{conv}\right) $ according to the counting operator $\mathcal{N}%
_\varepsilon =\varepsilon ^\mu \partial /\partial \varepsilon ^\mu \;$of eq.$%
\left( \ref{count-op}\right) \;$we get, to zeroth order in $\varepsilon ^\mu 
$,

\begin{equation}
\left( \frac{\partial \widetilde{\mathcal{S}}}{\partial g}\right)
_{\varepsilon =0}=b_{_{\widetilde{\mathcal{S}}}}\Xi _{\varepsilon =0}^{-1}\;,
\label{implic}
\end{equation}
with $\left( \partial \widetilde{\mathcal{S}}/\partial g\right)
_{\varepsilon =0}\;$given by

\begin{equation}
\left( \frac{\partial \widetilde{\mathcal{S}}}{\partial g}\right)
_{\varepsilon =0}=-\frac 2g\mathcal{S}_{TYM}+\frac{g\omega ^2}4\int
d^4x\;trB^{\mu \nu }B_{\mu \nu \;,}  \label{ds-dg}
\end{equation}
and $\mathcal{S}_{TYM}\;$being the original TYM action\ of eq.$\left( \ref
{tym}\right) .$\ However, it is very simple to check that the expression $%
\left( \ref{ds-dg}\right) $ can be rewritten as

\begin{eqnarray}
\left( g\frac{\partial \widetilde{\mathcal{S}}}{\partial g}\right)
_{\varepsilon =0} &=&\frac 2{3g^3}\varepsilon ^{\mu \nu \rho \tau }\mathcal{W%
}_\mu \mathcal{W}_\nu \mathcal{W}_\rho \mathcal{W}_\tau \int d^4x\;tr\left( 
\frac{\phi ^{2\;}}2\right) \;  \label{nontriv-ds-dg} \\
&&+b_{_{\widetilde{\mathcal{S}}}}\int d^4x\;tr\left( \phi D-\xi ^\mu \psi
_\mu \right) \;.  \nonumber
\end{eqnarray}
Therefore, according to the analysis of the previous section, it follows
that the term $\left( \partial \widetilde{\mathcal{S}}/\partial g\right)
_{\varepsilon =0}\;$ belongs to the integrated cohomology of $b_{_{%
\widetilde{\mathcal{S}}}}$. Equation $\left( \ref{conv}\right) \;$cannot
thus be satisfied, meaning that $\partial \widetilde{\mathcal{S}}/\partial g$
identifies indeed a nontrivial element of the cohomology of $\mathcal{B}_{%
\widetilde{\mathcal{S}}}$. We can conclude therefore that the symmetry
content of the topological Yang-Mills theory allows for a unique invariant
nontrivial counterterm whose most general expression can be written as

\begin{equation}
\Delta ^{count}=\mathcal{\varsigma }_g\frac{\partial \widetilde{\mathcal{S}}%
}{\partial g}+\mathcal{B}_{\widetilde{\mathcal{S}}}\Delta ^{-1\;},
\label{gen-count}
\end{equation}
$\mathcal{\varsigma }_g\;$being an arbitrary free parameter corresponding to
a possible renormalization of the gauge coupling constant $g$. The result $%
\left( \ref{gen-count}\right) $ is in complete agreement with that found in
ref. \textit{\cite{magg}} in the case of untwisted $N=2$ YM. This concludes
the algebraic renormalization analysis of Witten's TYM.

\begin{remark}
Observe that eqs.$\left( \ref{ds-dg}\right) $, $\left( \ref{nontriv-ds-dg}%
\right) $ imply that the expression of the original TYM action $\left( \ref
{tym}\right) $ can be rewritten as

\begin{eqnarray}
\mathcal{S}_{TYM} &=&-\left( \frac 1{3g^3}\varepsilon ^{\mu \nu \rho \tau }%
\mathcal{W}_\mu \mathcal{W}_\nu \mathcal{W}_\rho \mathcal{W}_\tau \int
d^4x\;tr\left( \frac{\phi ^{2\;}}2\right) \right) _{\varepsilon =\omega
=0}\;\;\;+\;b_{_{\widetilde{\mathcal{S}}}}(...)\;,  \nonumber
\label{sug-exp} \\
&&  \label{sug-exp}
\end{eqnarray}
This suggestive formula shows that the origin of the Witten's action can be
in fact traced back, modulo an irrelevant exact cocycle, to the invariant
polynomial $tr\phi ^2$.
\end{remark}

\begin{remark}
Let us also recall here that the explicit Feynman diagrams computation
yields a nonvanishing value for the renormalization constant $\mathcal{%
\varsigma }_g$, meaning that TYM (understood here as the twisted\textit{\ }%
version of\textit{\ }$N=2$\textit{\ }YM\textit{) }possesses a nonvanishing $%
\beta $ function for the gauge coupling $g$. The latter agrees with that of
the pure $N=2$ untwisted Yang-Mills \textit{\cite{beta}}. Moreover, it is
well known that the $\beta $ function of $N=2$\thinspace Yang-Mills theory
receives only one loop order contributions \textit{\cite{beta1}}. This
important result is commonly understood in terms of the nonrenormalization
theorem for the $U(1)$ axial anomaly which, due to supersymmetry, belongs to
the same supercurrent multiplet of the energy-momentum tensor. This implies
that there is a relationship between the coefficient of the axial anomaly
and the $\beta $ function, providing then a useful argument in order to
understand the absence of higher order corrections for the $N=2\;$gauge
Yang-Mills theory. On the other hand, the formula $\left( \ref{sug-exp}%
\right) $ shows that the TYM\ action $\mathcal{S}_{TYM}$ is directly related
to the invariant polynomial $tr\left( \phi ^2\right) $. It is known since
several years that in the $N=2$ susy gauge theories the Green's functions
with the insertion of composite operators of the kind of the invariant
polynomials $\mathcal{P}_n(\phi )$ of eq.$\left( \ref{b-cohomology}\right) $
display remarkable finiteness properties and can be computed exactely, even
when nonperturbative effects are taken into account \textit{\cite{akmrv}}.
It is natural therefore to expect that the finiteness properties of $%
tr\left( \phi ^2\right) \;$are at the origin of the absence of higher order
corrections for the gauge $\beta $ function of both twisted and untwisted $%
N=2$ gauge theories. In other words the relation $\left( \ref{nontriv-ds-dg}%
\right) $ could give us an alternative understanding of the
nonrenormalization theorem for the $N=2\;$gauge $\beta $ function. We shall
hope to report soon on this aspect in a more formal and detailed work.
\end{remark}

.\vspace{5mm}

{\Large \textbf{Acknowledgements}}

S.P. Sorella is grateful to the organizing commitee of the School for the
kind invitation.

\newpage 

\appendix

\section{Appendix}

\subsection{Susy conventions in euclidean space-time}

The supersymmetric conventions adopted here are those which can be found in
ref. \textit{\cite{th}}. For the matrices $(\sigma _\mu ,\;\overline{\sigma }%
_\mu )\;$we have

\begin{equation}
\left( \sigma ^\mu \right) _{\alpha \dot{\alpha}}=(1,\stackrel{\rightarrow }{%
i\sigma })\;,\;\;\;\;\left( \overline{\sigma }^\mu \right) _{\dot{\alpha}%
\alpha }=(1,-\stackrel{\rightarrow }{i\sigma })\;,  \label{sig-matrices}
\end{equation}
$\stackrel{\rightarrow }{\sigma }\;$being the Pauli matrices. As usual, the
spinor indices $(\alpha ,\dot{\alpha})$ are rised and lowered by means of
the antisymmetric tensors $\varepsilon _{\alpha \beta },\;\varepsilon _{\dot{%
\alpha}\dot{\beta}}$. The matrices $(\sigma _\mu ,\;\overline{\sigma }_\mu ) 
$ obey the following algebra

\begin{eqnarray}
\sigma _\mu \overline{\sigma }_\nu +\sigma _\nu \overline{\sigma }_\mu
&=&2g_{\mu \nu \;,}  \label{alg-sigma} \\
\overline{\sigma }_\mu \sigma _\nu +\overline{\sigma }_\nu \sigma _\mu
&=&2g_{\mu \nu \;,}  \nonumber
\end{eqnarray}
as well as the completeness relations 
\begin{equation}
\left( \sigma ^\mu \right) _{\alpha \dot{\alpha}}\left( \overline{\sigma }%
_\mu \right) ^{\dot{\beta}\beta }=2\delta _\alpha ^\beta \delta _{\dot{\alpha%
}}^{\dot{\beta}}\;,  \label{comp-rel}
\end{equation}
with $g_{\mu \nu }\;$the flat euclidean metric $g_{\mu \nu }=diag(+,+,+,+).$

For the antisymmetric matrices $\sigma ^{\mu \nu }$ and $\overline{\sigma }%
^{\mu \nu }\;$we have respectively

\begin{eqnarray}
\sigma ^{\mu \nu } &=&\frac 12(\sigma ^\mu \overline{\sigma }^\nu -\sigma
^\nu \overline{\sigma }^\mu )\;,\;\;\;\;\;\;\sigma ^{\mu \nu }=\widetilde{%
\sigma }^{\mu \nu }=\frac 12\varepsilon ^{\mu \nu \rho \sigma }\sigma _{\rho
\sigma }\;,  \label{def-smn} \\
\overline{\sigma }^{\mu \nu } &=&\frac 12(\overline{\sigma }^\mu \sigma ^\nu
-\overline{\sigma }^\nu \sigma ^\mu )\;,\;\;\;\;\;\;\overline{\sigma }^{\mu
\nu }=-\widetilde{\overline{\sigma }}^{\mu \nu }=-\frac 12\varepsilon ^{\mu
\nu \rho \sigma }\overline{\sigma }_{\rho \sigma }\;,  \nonumber
\end{eqnarray}
and 
\begin{eqnarray}
\sigma ^\mu \overline{\sigma }^\nu &=&g^{\mu \nu }+\sigma ^{\mu \nu }\;,
\label{dec-rel-s} \\
\overline{\sigma }^\mu \sigma ^\nu &=&g^{\mu \nu }+\overline{\sigma }^{\mu
\nu }\;.  \nonumber
\end{eqnarray}
The following useful relations hold

\begin{eqnarray}
\sigma ^{\mu \nu }\sigma ^\lambda &=&g^{\nu \lambda }\sigma ^\mu -g^{\mu
\lambda }\sigma ^\nu -\varepsilon ^{\mu \nu \lambda \rho }\sigma _\rho \;,
\label{fin-rel} \\
\overline{\sigma }^{\mu \nu }\overline{\sigma }^\lambda &=&g^{\nu \lambda }%
\overline{\sigma }^\mu -g^{\mu \lambda }\overline{\sigma }^\nu +\varepsilon
^{\mu \nu \lambda \rho }\overline{\sigma }_\rho \;,  \nonumber \\
\sigma ^\lambda \overline{\sigma }^{\mu \nu } &=&g^{\mu \lambda }\sigma ^\nu
-g^{\nu \lambda }\sigma ^\mu -\varepsilon ^{\mu \nu \lambda \rho }\sigma
_\rho \;,  \nonumber \\
\overline{\sigma }^\lambda \sigma ^{\mu \nu } &=&g^{\mu \lambda }\overline{%
\sigma }^\nu -g^{\nu \lambda }\overline{\sigma }^\mu +\varepsilon ^{\mu \nu
\lambda \rho }\overline{\sigma }_\rho \;.  \nonumber
\end{eqnarray}

\section{Appendix}

\subsection{Tensorial self-dual transformations}

As expected, the TYM action of eq.$\left( \ref{tym}\right) \;$is left
invariant by a set of nonlinear transformations whose generators $\delta
_{\mu \nu }\;$are self-dual, \textit{i.e. }$\delta _{\mu \nu }=\widetilde{%
\delta }_{\mu \nu }.\;\;$They read

\begin{eqnarray}
\delta _{\mu \nu }A_\sigma &=&-(\varepsilon _{\mu \nu \sigma \tau }\psi
^\tau +g_{\mu \sigma }\psi _\nu -g_{\nu \sigma }\psi _\mu )\;,
\label{tensor-transf} \\
\delta _{\mu \nu }\psi _\sigma &=&-(\varepsilon _{\mu \nu \sigma \tau
}D^\tau \phi +g_{\mu \sigma }D_\nu \phi -g_{\nu \sigma }D_\mu \phi )\;, 
\nonumber \\
\delta _{\mu \nu }\phi &=&0\;,  \nonumber \\
\delta _{\mu \nu }\overline{\phi } &=&8\chi _{\mu \nu }  \nonumber \\
\delta _{\mu \nu }\eta &=&-4F_{\mu \nu }^{+}  \nonumber \\
\delta _{\mu \nu }\chi _{\sigma \tau } &=&\frac 18(\varepsilon _{\mu \nu
\sigma \tau }+g_{\mu \sigma }g_{\nu \tau }-g_{\mu \tau }g_{\nu \sigma
})\left[ \phi ,\overline{\phi }\right] \;  \nonumber \\
&&+(F_{\mu \sigma }^{+}g_{\nu \tau }-F_{\nu \sigma }^{+}g_{\mu \tau }-F_{\mu
\tau }^{+}g_{\nu \sigma }+F_{\nu \tau }^{+}g_{\mu \sigma })  \nonumber \\
&&+(\varepsilon _{\mu \nu \sigma }^{\;\;\;\;\;\;\alpha }F_{\tau \alpha
}^{+}-\varepsilon _{\mu \nu \tau }^{\;\;\;\;\;\;\alpha }F_{\sigma \alpha
}^{+}+\varepsilon _{\sigma \tau \mu }^{\;\;\;\;\;\;\alpha }F_{\nu \alpha
}^{+}-\varepsilon _{\sigma \tau \nu }^{\;\;\;\;\;\;\alpha }F_{\mu \alpha
}^{+})\;.  \nonumber
\end{eqnarray}
The above transformations are checked to give rise, together with
transformations $\left( \ref{d-transf}\right) $ and $\left( \ref{vsusy}%
\right) $, to the complete twisted $N=2$ supersymmetric algebra of eqs.$%
\left( \ref{wz-talg-1}\right) $ and $\left( \ref{wz-talg-2}\right) $.

\begin{remark}
According to the algebraic set up of Sect.4, the tensor self-dual
transformations $\left( \ref{tensor-transf}\right) \;$can be easily encoded
in the Slavnov-Taylor identity $\left( \ref{n-tym-s-t}\right) \;$by means of
the introduction of a suitable constant tensor self-dual ghost. However, the
inclusion of the self-dual symmetry does not modify the previous results on
the renormalizability of TYM. The theory will remain anomaly free and will
admit a unique invariant nontrivial counterterm. As one can easily
understand, this is essentially due to the fact that the tensor
transformations $\left( \ref{tensor-transf}\right) $ do not actually act on
the scalar field $\phi $, so that they cannot modify the cohomology result
of eq.$\left( \ref{b-cohomology}\right) $.
\end{remark}

\end{document}